\documentclass[12pt,preprint]{aastex}

\usepackage{epstopdf}

\shorttitle{Alfv\'en wave heating}
\shortauthors{P. Antolin \and K. Shibata}

\begin{document}

\title{The role of torsional Alfv\'en waves in coronal heating}

\author{P. Antolin\altaffilmark{1,2}, K. Shibata\altaffilmark{1}}
\affil{\altaffilmark{1}Kwasan Observatory, Kyoto University,
    Yamashina, Kyoto, 607-8471, Japan}
\affil{\altaffilmark{2}The Institute of Theoretical Astrophysics, University of Oslo, P.O. Box 1029, Blindern, NO-0315 Oslo, Norway}
\email{antolin@kwasan.kyoto-u.ac.jp, shibata@kwasan.kyoto-u.ac.jp}
\altaffiltext{1}{Also at: Center of Mathematics for Applications, University of Oslo, P.O. Box 1053, Blindern, NO-0316, Oslo, Norway}

\begin{abstract}

In the context of coronal heating, among the zoo of MHD waves that exist in the solar atmosphere, Alfv\'en waves receive special attention. Indeed, these waves constitute an attractive heating agent due to their ability to carry over the many different layers of the solar atmosphere sufficient energy to heat and maintain a corona. However, due to their incompressible nature these waves need a mechanism such as mode conversion (leading to shock heating), phase mixing, resonant absorption or turbulent cascade in order to heat the plasma. Furthermore, their incompressibility makes their detection in the solar atmosphere very difficult. New observations with polarimetric, spectroscopic and imaging instruments such as those on board of the japanese satellite \textit{Hinode}, or the \textit{SST} or \textit{CoMP}, are bringing strong evidence for the existence of energetic Alfv\'en waves in the solar corona. In order to assess the role of Alfv\'en waves in coronal heating, in this work we model a magnetic flux tube being subject to Alfv\'en wave heating through the mode conversion mechanism. Using a 1.5-dimensional MHD code we carry out a parameter survey varying the magnetic flux tube geometry (length and expansion), the photospheric magnetic field, the photospheric velocity amplitudes and the nature of the waves (monochromatic or white noise spectrum). The regimes under which Alfv\'en wave heating produces hot and stable coronae is found to be rather narrow. Independently of the photospheric wave amplitude and magnetic field a corona can be produced and maintained only for long ($> 80$ Mm) and thick (area ratio between photosphere and corona $> 500$) loops. Above a critical value of the photospheric velocity amplitude (generally a few km s$^{-1}$) the corona can no longer be maintained over extended periods of time and collapses due to the large momentum of the waves. These results establish several constraints on Alfv\'en wave heating as a coronal heating mechanism, especially for active region loops.

\end{abstract}

\keywords{Sun: corona -- Sun: flares -- MHD -- waves}

\section{Introduction}

New observations from polarimetric, spectroscopic and imaging instruments are revealing a corona permeated with waves. Compressive modes such as the slow and fast magnetohydrodynamic (MHD) modes have a long chapter in observational history \citep[see reviews by ][]{Nakariakov_2005LRSP....2....3N, Banerjee_2007SoPh..246....3B, Ruderman_2009SSRv..tmp...54R, Taroyan_2009SSRv..tmp...24T}. Only recently, the development of high resolution instruments has brought strong evidence for the existence of the third MHD mode, the Alfv\'en mode, in the solar atmosphere. In the majority of the observational reports of Alfv\'en waves ambiguity exists and the waves can be interpreted as trapped modes, fast and slow kink waves. \citet{DePontieu_2007Sci...318.1574D} reported transversal displacements of spicules from observations in the Ca II H-line (3968 \AA) with a broadband filter of \textit{Hinode}/SOT. The wavelengths of these chromospheric agents were estimated to be longer than 20000 km, with periods between 100 and 500 s and speeds of at least 50-200 km s$^{-1}$. In the absence of a stable waveguide they interpreted the swaying of the spicules as a result of upward propagating Alfv\'en waves. However, it was pointed out by  \citet{Erdelyi_2007Sci...318.1572E} that these waves were likely to be kink oscillations, due to the displacement of the axis of symmetry of the flux tube caused by the later. Using the Coronal Multi-Channel Polarimeter (\textit{CoMP}) \citet{Tomczyk_2007Sci...317.1192T} analyzed properties of infrared coronal emission line Fe\,XIII (1074.7 nm) across a large field of view with short integration times. Waves propagating along magnetic field lines with ubiquitous quasi-periodic fluctuations in velocity with periods between 200 and 400 s (power peak at 5-min) and negligible intensity variations were reported. Wavelengths and phase speeds were estimated to be higher to 250 Mm and 1 Mm s$^{-1}$ respectively. In this work it was suggested that these waves were Alfv\'en waves probably being generated in the chromospheric network from mode conversion of p-modes propagating from the photosphere (hence explaining the peak in the power spectrum of the waves). On the other hand \citet{Erdelyi_2007Sci...318.1572E} and 
\citet{VanDoorsselaere_2008ApJ...676L..73V} argued that an interpretation in terms of fast kink waves was more appropriate due mainly to the observed collective behavior which should be absent in the case of Alfv\'en waves. Being compressible waves it was argued that kink waves would appear however incompressible in the corona for an instrument such as \textit{CoMP} due to the very small variation in intensity they produce.

The difficulty in detecting the Alfv\'en wave is due in part to its incompressible nature. Being only a transverse disturbance in the magnetic field, it is practically invisible to imaging instruments, unless the structure of propagation is displaced periodically from the line of sight \citep{Williams_2004ESASP.547..513W}. Also, due to the large wavelengths (a few megameters) and short periods (a few minutes) that may be involved in the corona, a large field of view, short integration time and proper resolution are needed for their detection in the corona. With polarimeters and spectrographs however Alfv\'en waves are more easily detected, as suggested by \citet{Erdelyi_2007Sci...318.1572E}. These waves propagate as torsional disturbances of the magnetic field, causing a periodic and spatially dependent spectral line broadening \citep{Zaqarashvili_2003AA...399L..15Z}. Up to date, the only observational report on torsional Alfv\'en waves is the recent work by \citet{Jess_2009Sci...323.1582J}, in which observations in H$\alpha$ of a bright point using the Solar Optical Universal Polarimeter (SOUP) of the \textit{SST} were reported. No variation in intensity nor line-of-sight velocity was detected and a periodic spectral broadening of the H$\alpha$ line was found, leaving Alfv\'en waves as the only possible interpretation for their results. The waves were reported to have periods in the 100-500 s interval (with maximum power in the 400-500 s range). By comparing H$\alpha$ continuum with H$\alpha$ core images they reported a canopy-like structure with magnetic fields expanding $\sim1300$ km in a vertical distance of $\sim1000$ km. This large expansion is crucial for Alfv\'en wave heating theories. For instance, assuming an Alv\'en speed of 10 km s$^{-1}$ in the upper photosphere/ lower chromosphere and a period of 100 s (a value in the lower range of periods reported in the works cited above) we obtain a wavelength of 1000 km for Alfv\'en waves propagating from the photosphere, matching the distance of large expansion of the flux tubes. Moreover, the height in the atmosphere where this takes place is the region of transition from high beta to low beta plasma. We thus have ideal conditions for mode conversion. Trapped modes will thus exchange their energy and a wide range of waves with different characteristics are likely to issue from this region. 

The large emphasis that has been put on the search for Alfv\'en waves in the solar corona is largely due to their connection with coronal heating. Theoretically, they can be easily generated in the photosphere by the constant turbulent convective motions, which inputs large amounts of energy into the waves \citep{Muller_1994AA...283..232M, Choudhuri_1993SoPh..143...49C}. Having magnetic tension as its restoring force the Alfv\'en wave is less affected by the large transition region gradients with respect to other modes. Also, when traveling along thin magnetic flux tubes they are cut-off free since they are not coupled to gravity (Musielak et al. 2007)\footnote{\citet{Verth_etal_aap_09} have pointed out however that the assertion made by \citet{Musielak_2007ApJ...659..650M} is valid only when the temperature in the flux tube does not differ from that of the external plasma. When this is not the case a cut-off frequency is introduced.}. Alfv\'en waves generated in the photosphere are thus able to carry sufficient energy into the corona to compensate the losses due to radiation and conduction, and, if given a suitable dissipation mechanism, heat the plasma to the high million degree coronal temperatures \citep{Uchida_1974SoPh...35..451U, Wentzel_1974SoPh...39..129W, Hollweg_1982SoPh...75...35H, Poedts_1989SoPh..123...83P, Ruderman_1997AA...320..305R, Kudoh_1999ApJ...514..493K} and power the solar wind \citep{Suzuki_2006JGRA..11106101S, Cranmer_2007ApJS..171..520C}.

Another possible generation mechanism for Alfv\'en waves is through magnetic reconnection. The amount of energy imparted to these waves during the reconnection process may depend on the location in the atmosphere of the event and is a subject of controversy. \citet{Parker_1991ApJ...372..719P} suggested a model in which 20~\% of the energy released by reconnection events in the solar corona is transfered as a form of Alfv\'en wave. \citet{Yokoyama_1998ESASP.421..215Y} studied the problem simulating reconnection in the corona, and found that less than 10~\% of the total released energy goes into Alfv\'en waves. This result is similar to the 2-D simulation results of photospheric reconnection by \citet{Takeuchi_2001ApJ...546L..73T}, in which it is shown that the energy flux carried by the slow magnetoacoustic waves is one order of magnitude higher that the energy flux carried by Alfv\'en waves. On the other hand, recent simulations by Kigure et al. (private communication) show that the fraction of Alfv\'en wave energy flux in the total released magnetic energy during reconnection in low $\beta$ plasmas may be significant (more than 50~\%). Since the observed ubiquitous intensity bursts (nanoflares) are thought to play an important role in the heating of the corona \citep{Hudson_1991SoPh..133..357H} and since they are generally assumed to be a signature of magnetic reconnection it is then crucial to determine the energy going into the Alfv\'en waves during the reconnection process. Moreover, it is possible that the observed nanoflares themselves are a consequence of Alfv\'en wave heating as porposed by\citet{Moriyasu_2004ApJ...601L.107M}. In that work the resulting intensity bursts producing the nanoflares are created by Alfv\'en waves by first converting to longitudinal modes which steepen into shocks and heat the plasma. This model was further developed by \citet{Antolin_2008ApJ...688..669A} (hereafter, Paper 1) and \citet{Antolin_2009arXiv0903.1766A}, where it was shown that the frequency of the resulting heating events from Alfv\'en waves followed a power law distribution. It was further shown that Alfv\'en wave heating could be differentiated from nanoflare-reconnection heating during observations through a series of signatures \citep[see also][]{Taroyan_2008IAUS..247..184T, Taroyan_2009SSRv..tmp...24T}. For instance, Alfv\'en waves lead to a dynamic, uniformly heated corona with steep power law indexes (issuing from statistics of heating events) while nanoflare-reconnection heating leads to lower dynamics (unless catastrophic cooling takes place in the case of footpoint concentrated heating) and shallow power laws.
 
The main problem faced by Alfv\'en wave heating is to find a suitable dissipation mechanism. Being an incompressible wave it must rely on a mechanism by which to convert the magnetic energy into heat. Several dissipation mechanisms have been proposed, such as parametric decay \citep{Goldstein_1978ApJ...219..700G, Terasawa_1986JGR....91.4171T}, mode conversion \citep{Hollweg_1982SoPh...75...35H, Kudoh_1999ApJ...514..493K, Moriyasu_2004ApJ...601L.107M}, phase mixing \citep{Heyvaerts_1983AA...117..220H, Ofman_2002ApJ...576L.153O}, or resonant absorption \citep{Ionson_1978ApJ...226..650I, Hollweg_1984ApJ...277..392H, Poedts_1989SoPh..123...83P, Erdelyi_1995AA...294..575E}. The main difficulty lies in dissipating sufficient amounts of energy in the correct time and space scales. For more discussion regarding this issue the reader can consult for instance \citet{Klimchuk_2006SoPh..234...41K, Erdelyi_2007AN....328..726E} and \citet{Aschwanden_2004psci.book.....A}.

Due to the observed increasing importance of Alfv\'en waves in the solar atmosphere in this work we address the subject of coronal heating by Alfv\'en waves in which mode conversion and parametric decay are taken as dissipation mechanisms. We concentrate on the heating of closed magnetic structures which populate the solar atmosphere, such as coronal loops, and analyze the efficiency of this heating model by carrying out a parametric space survey. We take a 1.5-dimensional model and we consider different photospheric drivers (a random driver creating a white noise spectrum, or a monochromatic driver in which several periods are tested), vary the photospheric magnetic field, the loop expansion from the photosphere to the corona and the loop length, and determine the regimes for which Alfv\'en wave heating plays an important role in coronal heating. The paper is organized as follows. In \S\ref{two} we set up the loop geometries and define the equations. As in Antolin et al. (2008) the loops are modeled with a 1.5-dimensional MHD code including thermal conduction and radiative cooling. In \S\ref{three} we analyze the effect of varying the parameters on the thermodynamic structure of the loop. Further discussion and conclusions are presented in \S\ref{four}.

\section{Alfv\'en wave model}\label{two}

\subsection{Parameter space}\label{model}

The parametric survey in this work involves parameters changing the geometry of the loop, such as the loop's length and its expansion from the photosphere to the corona. We consider lengths of 100, 80 and 60 Mm, leading to lengths of the corona of roughly 70, 50 and 20 Mm respectively, which are typical lengths for coronal loops. 

Due to the passage from gas pressure-dominated (high $\beta$) to magnetic pressure-dominated (low $\beta$) plasmas as we move to higher layers from the photosphere, magnetic flux tubes are expected to expand in the photosphere and chromosphere, forming the so-called ``magnetic canopy", and have a roughly constant cross-section area in the corona \citep{Aschwanden_2005ApJ...633..499A}. The loop expansion is a parameter influencing considerably the behavior of the waves as they propagate through the solar atmosphere. Apart from influencing the density stratification, it also introduces steep gradients in the Alfv\'en velocity due to the gradient in the magnetic field. The latter will in turn introduce different cut-offs for the waves propagating along the structure. Also, refraction of the fast MHD mode can easily occur, as fast MHD waves will always tend to follow the regions of higher $\beta$ plasma. Hence, expansion can determine not only the properties but also the nature of the waves that propagate along the magnetic field. Here we consider different expansions of the loop displaying a cross-sectional area ratio between the photosphere and the corona of 1000, 600 and 500. 

Apart from the previous geometrical parameters, we also consider different photospheric magnetic field strengths. Observations have shown the existence of kilo-gauss photospheric flux tubes all over the solar surface. Values have been reported ranging from below 1 kG to above 2 kG with a mean around 1.5 kG \citep{Solanki_1993SSRv...63....1S}. Here we survey different values for the photospheric magnetic field at the foopoint of the loop, namely, 2.3, 1.5 and 1 kG, and investigate the influence on the thermodynamics of the loop. 

Another parameter we consider concerns the photospheric driver generating the Alfv\'en waves.
So far the few observations of Alfv\'en waves in the solar atmosphere seem to indicate a broad spectrum range roughly between 100 and 500 s (see introduction) with a power peak close to the 5-min power peak of the p-modes. This may indicate that Alfv\'en waves share with the p-modes the photosphere as generation site. The waves could be a consequence of the constant buffeting of the magnetic field by convection, or possibly, photospheric magnetic reconnection. If, instead, Alfv\'en waves are primarily generated by magnetic reconnection higher up in the atmosphere it is difficult to explain the observed common power peak. A possible scenario  could be one in which p-modes trigger magnetic reconnection as in the recent simulations by \citet{Heggland_2009ApJ...702....1H}, and by some mechanism transfer the same periodicity to the waves. In this work we assume mainly that Alfv\'en waves are generated in the photosphere and have a white noise spectrum, which would issue from the random buffeting motion by convection (or maybe photospheric magnetic reconnection). However, we also study the influence of the wave period in the resulting corona. Hence we consider as well monochromatic wave drivers generating longitudinal modes with 10, 25, 50, 100, 150, 200 and 300 s period (as one sinusoidal torsional wave creates two longitudinal waves, the periods of the Alfv\'en waves are twice the periods stated here). This analysis was started in \citet{Antolin_2008ApJ...688..669A}, where it was shown that the most effective waves for coronal heating had periods between 100 and 150 s. Here we extend the analysis by studying in detail the hydrodynamic response of the loop to the different wave periods. The more realistic random photospheric driver is considered for the rest of the parameter space study.

The standard model of the loop to which we apply the different parameters is the same as the one considered in Paper 1. That is, a 100 Mm length loop with an apex-to-base area expansion of 1000, a photospheric magnetic field of 2.3 kG and in which the Alfv\'en waves are generated by a random photospheric driver creating a white noise frequency spectrum.

\subsection{MHD equations}

The magnetic flux tube is modeled with the local curvilinear coordinates ($s,\phi,r$) where $s$ measures distance along the most external poloidal magnetic field line, $\phi$ is the azimuthal angle measured around the rotation axis of the flux tube, and $r$ is the radius of the tube. We take the 1.5-D approximation,
\begin{equation}
\frac{\partial}{\partial\phi} = 0, \hspace{0.5cm} \frac{\partial}{\partial r} = 0, \hspace{0.5cm} v_{r}=0, \hspace{0.5cm} B_{r}=0,
\end{equation}
where $v_{r}$ and $B_{r}$ are, respectively, the radial components of the velocity and magnetic field in the magnetic flux tube. In the considered approximation, conservation of magnetic flux defines the value of the poloidal magnetic field as a function of $r$ alone, $B_{s}=B_{0}(r_{0}/r)^{2}$, where $B_{0}$ is the value of the magnetic field at the photosphere and $r_{0}=200$ km is the initial radius of the loop. In the photosphere the value of $\beta=8\pi p/(B_{s}^{2}+B_{\phi}^{2})$ (the ratio of gas to magnetic pressures) is unity. We assume an inviscid perfectly conducting fully ionized plasma. The effects of thermal conduction and radiative cooling are considered. 

The 1.5-D MHD equations are written as follows.
The mass conservation equation:
\begin{equation}\label{mass}
\frac{\partial\rho}{\partial t}+v_{s}\frac{\partial\rho}{\partial s}=
    -\rho B_{s}\frac{\partial}{\partial s}\left(\frac{v_{s}}{B_{s}}\right);
\end{equation}
the s-component of the momentum equation:
\begin{equation}\label{momentums}
\frac{\partial v_{s}}{\partial t}+v_{s}\frac{\partial v_{s}}{\partial s}=
    -\frac{1}{\rho}\frac{\partial p}{\partial s}-g_{s}+\frac{v_{\phi}^{2}}{r}\frac{\partial r}{\partial s} 
    -\frac{1}{4\pi \rho}\frac{B_{\phi}}{r}\frac{\partial}{\partial s}(r B_{\phi});
\end{equation}
the $\phi-$component of the momentum equation:
\begin{equation}\label{momentumfi}
    \frac{\partial (r v_{\phi})}{\partial t}+v_{s}\frac{\partial (r v_{\phi})}{\partial s} = \frac{B_{s}}{4\pi\rho}\frac{\partial}{\partial s}(r B_{\phi})+\mathcal{C}(t,s);
\end{equation}
the induction equation:
\begin{equation}\label{induction}
	\frac{\partial}{\partial t}\left(\frac{B_{\phi}}{r B_{s}}\right) 
	+ \frac{\partial}{\partial s}\left(\frac{B_{\phi}}{r B_{s}}v_{s}-\frac{v_{\phi}}{r}\right) = 0 ;
\end{equation}
and the energy equation:
\begin{equation}\label{energy}
\frac{\partial e}{\partial t}+v_{s}\frac{\partial e}{\partial s}=
    -(\gamma-1)e B_{s}\frac{\partial}{\partial s}\left(\frac{v_{s}}{B_{s}}\right)
    -\frac{R-\mathcal{S}}{\rho}+\frac{1}{\rho r^{2}}\frac{\partial}{\partial
    s}\left(r^{2}\kappa\frac{\partial T}{\partial s}\right);
\end{equation}
where 
\begin{equation}\label{glawinte}
        p=\rho\frac{k_{B}}{m}T, \hspace{0.5cm} e=\frac{1}{\gamma-1}\frac{p}{\rho}.
\end{equation}
In the above equations (\ref{mass})-(\ref{energy}) $\rho$, $p$, and $e$ are, respectively, density, pressure and internal energy; $v_{s}$ is the poloidal component of the velocity along the external magnetic field line; $v_{\phi}$ is the toroidal (azimuthal) component of the velocity; $B_{s}$ and $B_{\phi}$ are, respectively, the poloidal and toroidal components of the magnetic field; $k_{B}$ is the Boltzmann constant and $\gamma$ is the ratio of specific heats for a monatomic gas, taken to be 5/3. The $g_{s}$ is the effective gravity along the external poloidal magnetic field line and is given by
\begin{equation}\label{gravitys}
    g_{s}=g_{\odot}\cos\left(\frac{z}{L}\pi\right)\frac{dz}{ds},
\end{equation}
where $g_{\odot}=2.74\times10^{4}$ cm s$^{-2}$ is the gravity at the base of the loop, $z$ is the length along the central axis of the loop, and $L$ is the total length of the loop. The $\mathcal{C}(t,s)$ in equation (\ref{momentumfi}) is a term which simulates the torque motions in the photosphere. It is responsible for the generation of Alfv\'en waves. For random photospheric perturbations generating a white noise spectrum for the Alfv\'en waves we have
\begin{eqnarray}\label{torquerandom}
\mathcal{C}(t,s) = 2r(s)[\mbox{rand1}(t)-0.5]f\left\{\tanh\left(\frac{s-0.55H_{0}}{0.055H_{0}}\right)-1\right\}+\nonumber\\ + 2r(s)[\mbox{rand2}(t)-0.5]f\left\{\tanh\left(\frac{L-s-0.55H_{0}}{0.055H_{0}}\right)- 
1\right\},
\end{eqnarray}
where $H_{0}$ denotes the pressure scale height at $z=0$. We take $H_{0}=200$ km. The terms rand1$(t)$ and rand2$(t)$ are non correlated functions that output a number randomly distributed between 0 and 1, which changes in time, and $f$ is a parameter that determines the strength of the torque. For the case of the photospheric driver generating monochromatic Alfv\'en waves we have
\begin{eqnarray}\label{torquemono}
\mathcal{C}(t,s) = r(s)f\sin(2t\pi/\tau)\left\{\tanh\left(\frac{s-0.55H_{0}}{0.055H_{0}}\right)-1\right\}+\nonumber\\ + r(s)f\sin(2t\pi/\tau)\left\{\tanh\left(\frac{L-s-0.55H_{0}}{0.055H_{0}}\right)- 
1\right\},
\end{eqnarray}  
where $\tau$ is the period of the Alfv\'en waves. We consider waves with periods of 20, 50, 100, 200, 300, 400 and 600 s. As Alfv\'en waves, through mode conversion, generate longitudinal modes having half the period, we obtain longitudinal waves with periods of 10, 25, 50, 100, 150, 200 and 300 s. Since longitudinal waves are more easily detected during observations in this paper we prefer referring to the longitudinal waves that issue from the Alfv\'en waves. Hence the periods that are mentioned always refer to periods of longitudinal waves, unless stated otherwise. 

In equation (\ref{energy}), $\kappa=9\times10^{-7}$ erg s$^{-1}$ K$^{-1}$ cm$^{-1}$ is the Spitzer conductivity corresponding to a fully ionized plasma. Here $\mathcal{S}$ is a heating term maintaining the initial temperature distribution of the loop. The radiative
losses $R(T)$ are defined as 
\begin{equation}\label{opticalthin}
    R(T)=n_{e}n_{p}Q(T)=\frac{n^{2}}{4}Q(T)
\end{equation}
where $n=n_{e}+n_{p}$ is the total particle number density ($n_{e}$ and $n_{p}$ are, respectively, the electron and proton number densities, and we assume $n_{e}=n_{p}=\rho/m$ to satisfy plasma neutrality, with $m$ the proton mass) and $Q(T)$ is the radiative loss function for optically thin plasmas \citep{Landini_1990AAS...82..229L} which is approximated with analytical functions of the form $Q(T)=\chi T^{\gamma}$. We take the same approximation as in \citet[][; please refer to their Table 1]{Hori_1997ApJ...489..426H}. For temperatures below $4\times10^{4}$ K we assume that the plasma becomes optically thick. In this case, the radiative losses $R$ can be approximated by $R(\rho)=4.9\times10^{9}\rho$ \citep{Anderson_1989ApJ...336.1089A}. In equation (\ref{energy})  the heating term $\mathcal{S}$ has a constant non-zero value which is non-negligible only when the atmosphere becomes optically thick. Its purpose is mainly for maintaining the initial temperature distribution of the loop.

\subsection{Initial conditions and numerical code}

In the present model a sub-photospheric region is considered by adding 2 Mm at each footpoint of the loop, in which the radius of the loop is kept constant (hence keeping a constant magnetic flux). We take the origin $s=z=0$ as the top end of this region. The loop is assumed to follow hydrostatic pressure balance in the sub-photospheric region and in the photosphere up to a height of $4H_{0}=800$ km, where $H_{0}$ is the pressure scale height at $z=0$. The inclusion of the sub-photospheric region avoids unrealistic density oscillations due to the reflection of waves at the boundaries, thus avoiding any influence from the boundary conditions on the coronal dynamics. For the rest of the loop, density decreases as $\rho\propto h^{-4}$, where $h$ is the height from the base of the loop. This is based on the work by \citet{Shibata_1989ApJ...338..471S, Shibata_1989ApJ...345..584S} in which the results of 2D MHD simulations of emerging flux by Parker instability exhibit such pressure distribution. The initial temperature all along the loop is set at $T=10^{4}$ K. The density at the photosphere ($z=0$) is set at $\rho_{0}=2.53\times10^{-7}$ g cm$^{-3}$, and, correspondingly, the photospheric pressure is $p_{0}=2.09\times10^{5}$ dyn cm$^{-2}$. As the plasma $\beta$ parameter is chosen to be unity in the photosphere, the equipartition magnetic field strength $B_{eq}=\sqrt{8\pi p_{0}}$ corresponds to $B_{s,0}= 2.3$ kG. For loops having the later as the photospheric field strength, the corresponding value of the magnetic field at the top of the loop is then $B_{s,top}=2.3$, 3.8 and 4.6 G respectively for loops expansions of 1000, 600 and 500. 

In order to correctly resolve the dynamics in the chromosphere and transition region we take a fine spatial resolution in the numerical scheme by setting the grid size $\Delta s=0.025 H_{0}=5$ km up to a height of $\sim$16000 km. Then, the grid size is allowed to increase as $\Delta s_{i+1}=1.03 \Delta s_{i}$ until it reaches a size of 20 km. The size is then kept constant up to the apex of the loop. We take rigid wall boundary conditions at the photosphere.  The numerical schemes adopted are the CIP scheme \citep{Yabe_1991CoPhC..66..219Y} and the MOC-CT scheme \citep{Evans_1988ApJ...332..659E, Stone_1992ApJS...80..791S}. Refer to \citet{Kudoh_1998ApJ...508..186K} for details about the application of these schemes. The total time of the simulation is 568 minutes.

\section{Parametric space survey}\label{three}

In paper 1 it was shown that a corona heated by Alfv\'en waves is characterized by large dynamics. High speed plasma flows and large up and down shifts of the transition region due to the shocks are common features of this heating mechanism. In order to determine the average values of thermodynamic variables in the corona the location of the two transition regions are tracked down in time. This is carried out for all the figures displaying coronal quantities in the present work.

\subsection{Wave dissipation}\label{dissipation}

In the present model the linearly polarized Alfv\'en waves have basically only one mechanism available in order to dissipate their energy: mode conversion. Due to the large density stratification of the atmosphere Alfv\'en waves grow in amplitude and the problem easily becomes nonlinear. Alfv\'en waves can then convert a part of their energy to the longitudinal slow and fast modes during propagation. This conversion happens mainly while propagating from the photosphere to the chromosphere (passing from high-$\beta$ to low-$\beta$ plasma), although important conversion processes also happen in the corona, as discussed below. The resulting longitudinal modes subsequently steep into shocks and heat the plasma\footnote{It should be noted that being essentially a one dimensional simulation the wave vector is in the same direction as the magnetic field, and hence fast modes and Alfv\'en waves are degenerated from each other.}. This mode conversion mechanism has been shown to be an efficient dissipation mechanism for Alfv\'en waves by which the corona can be heated for both open and closed magnetic field regions \citep{Hollweg_1982SoPh...75...35H, Kudoh_1999ApJ...514..493K, Moriyasu_2004ApJ...601L.107M,  Antolin_2008ApJ...688..669A} and also for the generation of the fast and slow solar wind \citep{Suzuki_2006JGRA..11106101S}.  Another dissipation mechanism which may be playing an important role in our model, especially when considering the monochromatic wave driver in the photosphere, is through resonant damping. In this heating scenario closed magnetic flux structures like loops serve as resonant cavities for Alfv\'en waves. For certain wave frequencies the loop can resonate thus allowing a high transmission of these waves into the corona. This heating mechanism is further discussed in section \ref{driver}. For the mode conversion mechanism the density fluctuation is crucial for the efficient dissipation of the waves. As was shown in Paper 1 and other related works \citep[see for instance][]{Suzuki_2006JGRA..11106101S} the coronae issuing from Alfv\'en wave heating characterize by showing large density fluctuations. Mode conversion takes place wherever the sound and Alfv\'en speed have similar values (plasma $\beta\sim1$ region), which is mainly the chromosphere, but mainly when the nonlinear effects are large. Nonlinear effects can be quantified by the ratio of azimuthal to longitudinal magnetic field (or velocities) $B_{\phi}/B_{s}$. As shown in Paper 1, this quantity takes values higher than 0.5 sporadically and ubiquitously in the corona due to the large density fluctuation caused by the longitudinal shocks. The large density fluctuation throughout the loop also causes the deformation of the Alfv\'en wave shape, which causes small-scale fluctuations of the magnetic pressure and leads to the generation of slow modes\footnote{The mean Alfv\'en speed in the corona obtained from heating by 100 s period waves in our model is 300 km s$^{-1}$. This gives a wavelength on the order of one third of the loop size. Such long wavelengths and the large density fluctuations lead to the deformation of the wave shape and the subsequent generation of slow modes.} \citep{Moore_1991ApJ...378..347M}. Another factor playing an important role in the wave dissipation is the wave-to-wave interaction. In this closed magnetic field model we have waves being generated at both footpoints, and also reflection of waves at both transition regions. We thus have collision of waves and shocks at all times in the corona, which greatly enhances nonlinearity and shock strengths. This intermittent mode conversion in the corona leads to a bursty intensity profile similar to the usual X-ray intensity profiles of the solar corona (as those shown by \textit{Hinode}/XRT for instance). This has led \citet{Moriyasu_2004ApJ...601L.107M} to propose the idea of nanoflares being the signatures of this nonlinear heating mechanism from Alfv\'en waves. It has also been shown from statistics of the resulting heating events that a power law distribution of the intensities issues with an index steeper than 2 (cf. Paper 1), a scenario in which the bulk of the heating comes from the small nanoflare-like heating events \citep{Hudson_1991SoPh..133..357H}.

\subsection{General characteristics}\label{general}

Figs.~\ref{fig3}, \ref{fig5}, \ref{fig6} and \ref{fig7} show the average in time of several quantities describing the thermodynamic state of the corona with respect to the photospheric rms azimuthal velocity field created by the torsional motions at the footpoints of the loop for each case of our parameter study. In panels a of Figs.~\ref{fig3}, \ref{fig5}, \ref{fig6} and \ref{fig7} we see the expected general tendency of the mean coronal temperature to increase with the amplitude of the photospheric velocity perturbation. This is expected since higher photospheric transversal velocities mean a higher energy flux into the loop (the converse is not always true as we will see below), as shown by panels c in Figs.~\ref{fig3}, \ref{fig5}, \ref{fig6} and \ref{fig7}, which show the variation of the mean rms photospheric magnetic energy $\langle B_{\phi, ph}^{2}/8\pi\rangle^{1/2}$ with the photospheric velocity amplitude. Higher energy fluxes from the waves generate stronger longitudinal shocks leading to a higher heating rate as long as conditions for mode conversion are met (see section\ref{dissipation}). This is shown in the e panels corresponding to the same figures, where the volumetric heating rate $H$ is plotted (below we explain how we calculate this quantity). We see that conditions leading to long term average temperatures above 1 MK are not easily met. Only large expansions (above area ratio of 600 between photosphere and corona) together with long lengths (above 100 Mm) provide the necessary conditions. Also, the photospheric velocity amplitudes must be large enough. In the present case coronae with average temperatures (over the entire simulation time) above 1 MK are obtained for $\langle v_{\phi, ph}^{2}\rangle^{1/2}>1.6$ km s$^{-1}$. On shorter timescales, temperatures in the corona can easily reach nanoflare-like temperatures above 1 MK, as shown by panels g, which shows the maximum temperatures that can be reached with Alfv\'en wave heating. Maximum temperatures show the same dependence with the photospheric velocity field as for the mean coronal temperature. Maximum temperatures between 2 and 5 MK are found for long period waves (defined here as waves with periods above 100 s). This is further discussed in section \ref{driver}. As was found in Paper 1 the most efficient wave periods for the heating are between 100 and 150 s. The 300 s waves also offer an interesting heating scenario. Their high power (as seen in the c panel of Fig.~\ref{fig3}) combines with resonant damping to give the highest temperatures in the model. This case is discussed in detail in section~\ref{driver}.

As the strengths of the shocks increase with the amplitude of the photospheric perturbations the mass flux into the corona increases due to the higher magnetic pressure from the waves, thus increasing the density of the corona (b panels). The Alfv\'en speed consequently decreases on the one hand, while the sound speed increases on the other due to the higher temperatures, leading to a high increase of the plasma $\beta$ parameter (c panels), which is the ratio of gas to magnetic pressures $\beta=8\pi p/(B_{s}^{2}+B_{\phi}^{2})$. Due to the strong shocks permeating the corona, the plasma $\beta$ parameter takes values above 0.1 sporadically and ubiquitously in the corona, contributing to mode conversion processes. This behavior is also shared by the quantity $v_{\phi}/v_{A}$, which measures the nonlinear effects and therefore is the main contributor for mode conversion. This is crucial for the efficient dissipation of the waves in the corona, which therefore gets uniformly heated, as discussed in section \ref{dissipation}. The average value in time of this quantity along the loop is plotted in Fig.~\ref{fig2}. Panel a corresponds to our base model: a loop of length 100 Mm, photospheric magnetic field of 2.3 kG and area expansion of 1000 between the photosphere and the corona, in which a white noise spectrum of Alfv\'en waves is generated at the footpoints. Panels b, c and d correspond to variations of the base model: with a photospheric magnetic field of 1 kG (panel b), an area expansion of 500 (panel c), and a length of 60 Mm (panel d). For each model different values of the rms photospheric velocity field are plotted. The solid line corresponds to the value of $\langle v_{\phi, ph}^{2}\rangle^{1/2}$ for which the highest temperatures are obtained in each model. As a general rule in all our models a high value of the ratio $v_{\phi}/v_{A}$ in the corona ($\gtrsim$ 0.3), or correspondingly in the photosphere ($\gtrsim$ 0.06\footnote{The photosphere is pointed out in the figure by the location close to the footpoints where $|v_{\phi}|/v_{A}$} acquires its minimum value. In the sub-photospheric region the ratio increases again due to the low Alfv\'en velocity), is necessary in order to obtain a million degree corona. For higher amplitudes of the rms photospheric velocity field (dashed lines) the nonlinear effects are reduced (except for the case of the short loop with 60 Mm, which we will discuss in more detail in section \ref{length}) and the loop enters a regime in which the corona cannot be maintained over long periods of time. This regime is discussed in section \ref{collapse}. 

\begin{figure}
\epsscale{.55}
\plotone{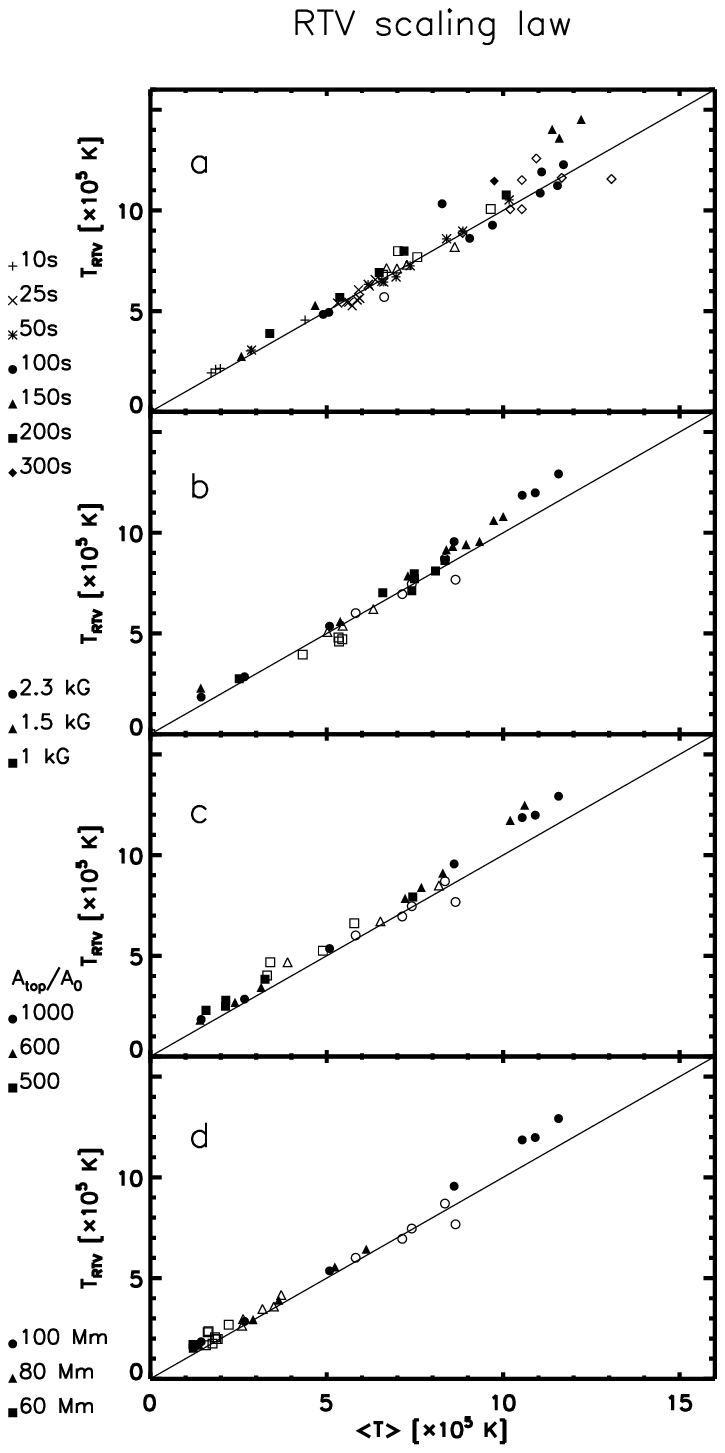}
\caption{Temperature predicted by the RTV scaling law (see section \ref{general}) with respect to the obtained average temperature of the corona for the case of Alfv\'en wave heating with monochromatic waves (panel a), different photospheric magnetic fields (panel b), different loop expansions (panel c), and different loop lengths (panel d). The values and corresponding symbols for each parameter are on the lower left side of each panel.  \label{fig1}}
\end{figure}

\begin{figure}
\epsscale{.47}
\plotone{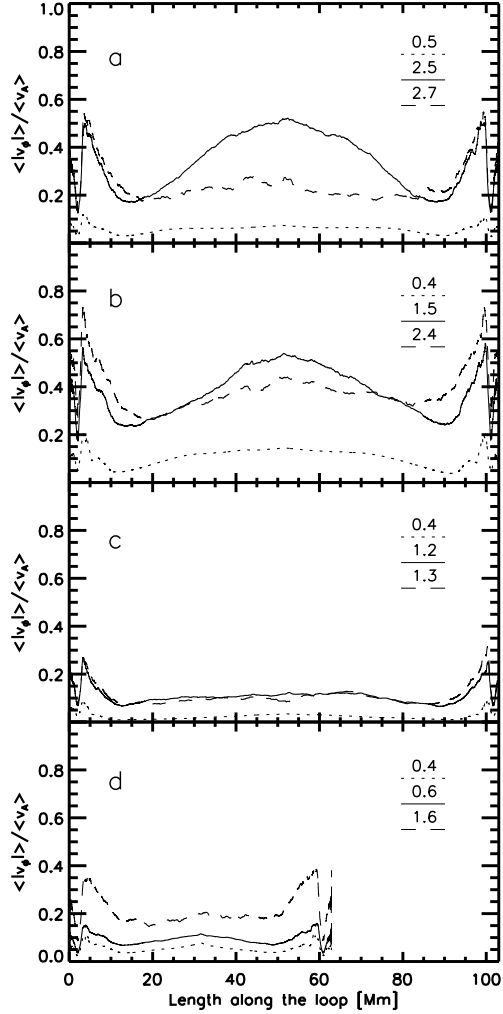}
\caption{Nonlinear effects quantified through the ratio of the azimuthal velocity to the Alfv\'en velocity, $\langle|v_{\phi}|\rangle/\langle v_{A}\rangle$. Here we plot the average value in time of this quantity along the loop for different models. Panel a: base model, a loop of length 100 Mm, photospheric magnetic field of 2.3 kG and area expansion of 1000 between photosphere and corona. Panels b, c and d: Same as base model but with a photospheric magnetic field of 1kG, with an area expansion of 500, with a length of 60 Mm, respectively. For each model three cases with different photospheric rms azimuthal velocity fields are plotted, the values of which are written in the upper right corner of each panel. The dotted lines correspond to a low value of $\langle v_{\phi}^{2}\rangle^{1/2}$ (leading to low coronal temperatures). The solid line corresponds to the value of $\langle v_{\phi}^{2}\rangle^{1/2}$ for which the highest temperatures are found. The dashed line corresponds to a value for which the corona collapses during the simulation (see section \ref{collapse}). \label{fig2}}
\end{figure}

The uniform heating of the corona can be quantified by comparing the obtained temperatures with the predicted coronal temperatures based on the RTV scaling law \citep{Rosner_1978ApJ...220..643R}. This is demonstrated in Fig.~\ref{fig1}. Panels a, b, c and d display the mean temperature in the corona with respect to the predicted temperature from the RTV scaling law, and correspond, respectively, to the photospheric driver study, the photospheric magnetic field, the loop expansion study and the loop length study. It can be noticed that in general the RTV scaling law fits remarkably well with the obtained coronal values, mainly for short period waves ($< 100$ s, as discussed in section \ref{driver}) and low photospheric velocity amplitudes. The error increase with photospheric velocity amplitude is due to the stronger shocks which input large amounts of chromospheric material in the corona (creating high spicule-like events) and therefore disrupts the local energy balance. Since the RTV scaling law is found to be well satisfied we estimate from it the volumetric heating rate $H$. As shown in \citet{Rosner_1978ApJ...220..643R}, we have:
\begin{equation}\label{hrtv}
H=9.8\times10^{4}p^{7/6}L^{-5/6}.
\end{equation}
The estimations resulting from this equation are plotted in the e panels of Figs.~\ref{fig3}, \ref{fig5}, \ref{fig6} and \ref{fig7}. 

Despite the higher momentum of the waves as the amplitudes of the photospheric driver increase the upward shifts of the transition region (spicule-like events) remain roughly constant. This has been previously found by \citet{Hollweg_1982SoPh...75...35H}. As explained in that paper for the case of open flux tubes, given a high amplitude shock we will have a faster upward rise of the transition region and thus also a longer duration for the following shock to reach the transition region, hence giving the latter more time to recede from gravity to lower heights. In the present case of closed flux tubes we have found the same behavior, thus producing a mean length of the corona that is roughly constant as seen in the f panels. However, when the photospheric velocity amplitude exceeds a critical value this is no longer the case. Indeed, a sharp decrease of the coronal length can be noticed when the photospheric azimuthal velocities exceed a certain critical value which depends mostly on the geometrical parameters of the loop (length and expansion). The critical photospheric velocity decreases the longer the periods of the waves as discussed in section \ref{driver}. For photospheric velocities higher than the critical velocity the corona enters an unstable regime in which the temperature, density (and consequently the plasma $\beta$ parameter) stop increasing. These cases are marked with blank symbols in all panels. We will now discuss this corona collapse regime.

\subsubsection{Corona collapse}\label{collapse}

In order to create a corona wave heating coming from mode conversion of Alfv\'en waves (cf. section \ref{dissipation}) has to exceed the losses due to radiation and (downward) conduction. The rapidity with which this energy balance is reached and a corona forms will depend on many parameters, such as the initial conditions as well as geometrical aspects (length and expansion of the loop for instance). In panels h of Figs.~\ref{fig3}, \ref{fig5}, \ref{fig6} and \ref{fig7} we show the time span of the corona over the entire simulation time (568 min). We can see that for each set of the considered loop parameters there is a range of photospheric velocity amplitudes for which the energy balance in the corona is maintained over most of the simulation, i.e. for which the corona is stable. The time span of the corona in this region is roughly constant. However there is a critical velocity above which the stability of the corona is compromised and the time span drops off sharply. These cases are marked with blank symbols in all panels. For these cases the corona can break down at some point and reform at a later moment, or not be reformed at all. More specifically, we have defined a stable corona as a corona which lasts for more than 60~\% of the total simulation time, which corresponds roughly to 5.7 hours. The value of the photospheric magnetic field (and hence the value of the background magnetic field) does not seem to influence the critical photospheric velocity at which the instability sets in, as seen in the h panel of Fig.~\ref{fig5}. For all the values considered for the photospheric magnetic field the corona gets unstable for $\langle v_{\phi, ph}^{2}\rangle^{1/2}\gtrsim2.2$ km s$^{-1}$ (see section \ref{mag}). Loops with thinner area cross-sections and shorter lengths have much smaller stable regimes (see sections \ref{expansion} and \ref{length}). Unstable coronae are seen to happen only for wave periods above 100 s. The critical velocity amplitude above which instability sets in is also dependent on wave period. The longer the period the lower the critical velocity. For instance, for 100 and 200 s waves we find unstable coronae already at $2.8$ km s$^{-1}$ and $1.8$ km s$^{-1}$ respectively.

The unstable regime can be understood partly by analyzing the behavior of the volumetric heating rate. As shown by the e panels of Figs.~\ref{fig3}, \ref{fig5}, \ref{fig6} and \ref{fig7} this quantity reaches a maximum for values of the photospheric velocity field which are generally lower than the critical velocity for which the corona collapses. The volumetric heating rate is then either constant or slightly decreases. Correspondingly, the mean temperature and density in the loop reach maximum values after which they are either constant or decrease. The presence of these maxima can be understood by means of the energy balance in the loop, which can be approximated in the following way:
\begin{equation}
H=\frac{n^{2}}{4}Q(T) + \kappa_{0}\frac{T^{7/2}}{L^{2}},
\end{equation}
where the first term on the right corresponds to the radiative losses, and the second term corresponds to thermal conduction. For values of the photospheric velocity field lower than the critical value we have a constant volumetric heating rate $H_{max}$, temperatures corresponding to an optically thin cooling which can be approximated by $Q(T)=\chi T^{-1/2}$ with $\chi= 10^{-18.48}$ \citep[we take the same approximation for temperatures above $10^{5.1}$ as in][]{Rosner_1978ApJ...220..643R} and a roughly constant length of the corona $L$. The maximum value of the volumetric heating rate then naturally sets a maximum value for the density and temperature. We have:
\begin{equation}
n=[\frac{4}{\chi}T^{1/2}(H_{max}-\kappa_{0}\frac{T^{7/2}}{L^{2}})]^{1/2}
\end{equation}
This equation defines a maximum density value $n_{max}=[\frac{7}{2\chi}H_{max}T_{m}^{1/2}]^{1/2}$ for a temperature $T_{m}=(\frac{H_{max}L^{2}}{8\kappa_{0}})^{2/7}$. Replacing by numerical values: $H_{max}\sim3\times10^{-5}$ erg cm$^{-3}$ s$^{-1}$, $L\sim8\times10^{9}$ cm, and $\kappa_{0}\sim10^{-6}$ erg s$^{-1}$ K$^{-1}$ cm$^{-1}$, we obtain a maximum density of $n_{max}\sim5\times10^{8}$ cm$^{-3}$ for a temperature of $T_{m}\sim6\times10^{5}$ K, which roughly match the critical temperature and density values we obtain.

Fig.~\ref{fig2} shows that nonlinear effects are reduced in the unstable regime. Hence, mode conversion is reduced leading to a saturation of the volumetric heating rate. The saturation is also due to the height of the spicules. As explained in the previous section, the height of the transition region is roughly unaffected by the amplitude of the photospheric driver, as long as this one does not exceed the critical value. When the critical amplitude is exceeded spicules become much longer due to the decreasing effective gravity along the loop. The cooling effect on the corona is therefore large and the volumetric heating rate from the waves is saturated. Correspondingly, for values of the photospheric velocity close to the critical velocity the energy balance along the loop is not well maintained, and the error from the RTV scaling law increases, as seen in Fig.~\ref{fig1}. This results in the collapse of the corona. As stated previously, long period waves have shorter critical velocity than short period waves. This is due to the stronger and less frequent shocks obtained with longer periods as discussed in section \ref{driver}. For 300 s waves the obtained coronae are unstable even with low amplitude photospheric velocities (except for only one case). Long period waves can carry large amounts of chromospheric material into the corona, increasing their density as shown by panel b of Fig.~\ref{fig3}. Since the energy flux into the corona is constant in time the heating per unit mass decreases in time, an effect which is stronger for long periods due to the lower number of shocks in the corona at a given time. For low amplitudes of the long period waves (that is, as long as the cooling effect from the spicules is not important) the coronae reach a critical state in which they are susceptible to a thermal instability, by which the corona rapidly cools down (in the order of minutes) and the loop is evacuated. A similar cooling event is obtained for loops which are heated towards their footpoints \citep{Antiochos_1999ApJ...512..985A, Muller_2003AA...411..605M, Mendozabriceno_2005ApJ...624.1080M}, a phenomenon termed ``catastrophic cooling'', which has been proposed as the physical explanation for coronal rain. The cyclic pattern of heating and cooling in the loops \citep[``limit cycles", as termed by][]{Muller_2003AA...411..605M} is only seen for low amplitudes of the monochromatic driver generating long period waves. In all other cases, for instance for thinner and shorter loops or for high amplitudes of the random photospheric driver, the collapse of the corona that is obtained is of a different nature than that of catastrophic cooling. In these cases the collapse of the corona is not due to a local increase of the radiative losses in the corona from the formation of a cool condensation as in the case of catastrophic cooling. Instead, the long spicules shorten dramatically the length of the corona, disrupting the local energy balance, and no condensation is formed. Whether Alfv\'en waves can reproduce coronal rain, as catastrophic cooling does, despite the uniform heating they produce is an interesting question which will be further investigated in a following paper. We will now discuss individual features of each set of parameters.

\subsection{Individual characteristics}\label{individual}

\subsubsection{Photospheric driver}\label{driver}

The loop model in this section consists on a 100 Mm length loop with an expansion factor of 1000 and a photospheric magnetic field of 2.3 kG, and in which monochromatic Alfv\'en waves are generated at both footpoints.

Panels a to h in Fig.~\ref{fig3} display the characteristics of the coronae produced by monochromatic slow and fast MHD waves resulting from the mode conversion of Alfv\'en waves. We can see that short period ($\leqslant50$ s) and long period ($\geqslant100$ s) waves produce very different coronae. Long period waves produce hotter and denser coronae than short period waves as shown by panels a and b. Coronae produced by long period waves are subject however to the unstable regime discussed in section~\ref{collapse} when the photospheric velocity amplitudes are high, as can be seen in panel h, and by the blank symbols in all the panels. For wave periods below 150 s, for the same velocity amplitude we obtain higher coronal temperatures the longer the period, and we find temperatures above 1 MK for velocities $\langle v_{\phi}^{2}\rangle^{1/2}$ in the range $0.9 - 1.6$ km s$^{-1}$ for 100 and 150 s period waves. Waves with lower periods (high frequency) are rapidly damped when propagating through the lower atmospheric layers. On the other hand, since longer periods imply longer wavelengths the number of shocks heating the corona at a given time is reduced and the corona is less uniformly heated. This is shown in panel a of Fig.~\ref{fig1}, where the match with the RTV scaling law is plotted \citep{Rosner_1978ApJ...220..643R}. Long period waves seem to suffer also from a period cut-off, reducing their efficiency for heating the solar corona. \cite{Musielak_2007ApJ...659..650M} have shown that torsional Alfv\'en waves have cut-off free propagation for the case of thin magnetic flux tubes, and thus can carry large amounts of energy into the corona. \citet{Verth_etal_aap_09} have pointed out however that the assertion made by \citet{Musielak_2007ApJ...659..650M} is valid only when the temperature in the flux tube does not differ from that of the external plasma. In the present case the flux tube presents velocity gradients (which issue from the density stratification and expansion of the magnetic flux tube) which introduce a period cut-off. \cite{Routh_2007SoPh..246..133R} have considered this problem analytically and shown that the cut-off period for torsional Alfv\'en waves is local (as opposed to global, where coupling to gravity or density stratification exists), introduced by gradients of the characteristic wave velocities. Caution should be taken however when applying the results of the previous paper, since it contains an approximation which may not be valid for thick flux tubes, as pointed out by \citet{Verth_etal_aap_09}. However, their main result concerning the existence of the cut-off period should remain valid. Hence, we can expect the cut-off period to vary with height in the same way as the velocity of the torsional waves $v_{A}$ varies. In view of the questionable approximation taken in their paper, we can take their reported cut-off period as an approximation. In the case of propagation through an isothermal medium they have shown that the period cut-off is given by

\begin{equation}\label{cutoff}
P_{cut}(s)=\frac{mh}{v_{A}(s)},
\end{equation}
where they approximate $v_{A}(s)=v_{A,0}\exp(s/mh)$, with $s$ the propagating distance, $h$ the pressure scale height and $m$ is a positive scaling factor. According to this result the propagation of Alfv\'en waves into the corona in our model should be significantly reduced. In our model the exponential increase in the Alfv\'en velocity happens in the upper chromosphere. In that region a typical value for the mean sound speed is $c_{s}=15$ km s$^{-1}$, giving a pressure scale height of $h\sim500$ km, and the mean Alv\'en velocity is $v_{A}=10$ km s$^{-1}$. Hence we get a cut-off period of $\sim50$ s for the torsional Alfv\'en waves, so 25 s for the longitudinal waves. This implies that effects from the cut-off period should be present in most of the cases considered. Since propagation of the Alfv\'en waves do happen (at least to heights where mode conversion occurs) resulting in the formation of a corona, the discrepancy could be due to the highly nonlinear effects in this model (and also to the questionable approximation of the paper). The importance of the nonlinear effects for Alfv\'en wave heating is thus emphasized. Apart from the cut-off period for Alfv\'en waves the normal cut-off period for longitudinal waves is also present. In our model, the value of this period cut-off is close to 200 s. Hence, a large fraction of longitudinal waves with periods larger than 200 s issuing from mode conversion occurring below the transition region won't be able to propagate into the corona. This may explain the low heating efficiency from the 200 s period waves. For the 300 s waves however, we still can have a hot corona as seen in the a panel. We will discuss why this is so further ahead. 

Long period waves can create very hot heating events close to 5 MK as shown in panel g, accounting for the observed nanoflares with instruments such as \textit{Hinode}/XRT \citep{Golub_2007SoPh..243...63G}. Despite the fact that short period waves do not contribute substantially to the hot temperatures of the corona they do have an important effect for the thermal stability of the loop. As previously stated the coronae obtained with short period waves are more uniform due to the higher number of shocks permeating the corona. The large cool mass upflows into the corona due to the waves are reduced, resulting in longer lengths and time spans of the corona (panels f and h).

In order to better understand these results, we plot in panel c of Fig.~\ref{fig3} the mean azimuthal magnetic energy $B_{\phi}^2/8\pi$ at the photosphere with respect to the photospheric velocity amplitude  for each period. $B_{\phi}$ is a measure of the twist exerted at the footpoints of the loop, which increases with the period of the wave. Indeed, we can see that for a same input of magnetic energy, there is a shift towards lower amplitudes of the photospheric velocity field the longer the period of the wave. Hence, it is to be expected that longer period waves produce hotter coronae as shown by panel a. Also, larger twists imply higher wave pressure and mass flux into the corona, thus explaining the higher densities obtained with long period waves. The 300 s waves exhibit a considerable shift: with a photospheric velocity amplitude as low as $0.25$ km s$^{-1}$, the corresponding magnetic energy input is as high as for the 100 s period waves with an amplitude of $\langle v_{\phi}^{2}\rangle^{1/2}=1.3$ km s$^{-1}$ (namely, $\sim1.8\times10^{4}$ erg cm). This explains why we obtain the same coronal temperatures for both cases (panel a). 

The large difference that can be seen in the photospheric azimuthal magnetic energy (panel c), and in general in all panels of Fig.~\ref{fig3} between the 300 s waves case and the other periods points to the presence of an additional heating mechanism. Indeed, apart from mode conversion there is another heating mechanism which seems to be playing an important role, especially when considering a monochromatic wave driver. Closed magnetic structures such as loops may act as resonant cavities for the Alfv\'en waves, which can suffer from reflection due to the steep gradients of the Alfv\'en velocity and be trapped in the corona. Resonances can increase the transmission coefficient from the waves, thus allowing large energy fluxes into the corona. Heating through resonances was first considered by \citet{Ionson_1978ApJ...226..650I}. 

In Fig.~\ref{fig4} we plot the map of the power spectrum at each point along the first 20 Mm of a loop with 300 s monochromatic waves for the longitudinal velocity component $v_{s}$ corresponding to the longitudinal modes, and azimuthal velocity component $v_{\phi}$ corresponding to the Alfv\'en waves. In this case the rms photospheric velocity amplitude is $<v_{\phi}^{2}>^{1/2}\sim 0.75$ km s$^{-1}$, and we can see a clear resonance pattern from the corona down to the chromosphere. In the longitudinal velocity map, the fundamental mode at 300 s and the first (150 s), second (100 s) and third (75 s) harmonics coming from the mode conversion of the Alfv\'en waves can be clearly seen. However, in the azimuthal velocity map the fundamental mode at 600 s and only the even harmonics (second, fourth, and so on) can be seen. Since we have set the monochromatic driver to generate waves that are in phase, the absence of the odd harmonics is simply due to a destructive interference between the Alfv\'en waves generated at both footpoints. Nevertheless, since mode conversion can happen in a time scale of minutes, before the disappearance of the odd harmonics mode conversion from these waves occurs resulting in the odd harmonics of the longitudinal velocity map. 

\citet{Hollweg_1984ApJ...277..392H} was the first to consider resonant damping in the case of a loop, where the later is approximated by a three-layer model in which the Alfv\'en velocity increases exponentially in the boundary layers (chromosphere and transition region) and is constant in the corona.  \citet{Hollweg_1984ApJ...277..392H} showed that when a wave is generated at one footpoint the transmission coefficient into the corona can be written
\begin{equation}
T\approx \left(1+\left(\frac{v_{A,c}\sin(kL_{c})}{2\pi h \omega}\right)^{2}\right)^{-1},
\end{equation}
where $v_{A,c}$ is the Alfv\'en speed in the corona, $L_{c}$ is the length of the corona, $h$ is the pressure scale height, $\omega$ is the frequency of the wave and $k=\omega/v_{A,c}$ is the wavenumber. The coronal resonances are excited for the periods
\begin{equation}\label{resonalf}
\tau_{res}=\frac{2L_{c}}{m v_{A,c}},
\end{equation}
where $m=1,2,3...$. At these periods the transmission coefficient has its maximum value. Eq.~\ref{resonalf} is simply the travel time for the Alfv\'en waves back and forth between the boundaries of the resonant cavity, and hence denotes the normal oscillating modes for the Alfv\'en waves. In that work the height of the transition region was however not considered, which is important in our model. Indeed, we have seen that the height of the spicules can be large. Furthermore, we also consider a sub-photospheric region at both footpoints. We can estimate the resonant periods in our model by adding the travel time of the waves in the regions below the transition regions. A more precise estimation of the resonant periods would take us too far from our present parameter study, and thus is left as future work. We have:
\begin{equation}\label{resonalf2}
\tau_{res}=2\frac{2L_{fp}}{m v_{A,fp}}+\frac{2L_{c}}{m v_{A,c}},
\end{equation}
where $L_{fp}$ is the length of the region below the transition region, and $v_{A,fp}$ is the mean Alfv\'en velocity in that region (all quantities denote averages over time). Inserting mean values of the 300 s waves case where resonance is detected: $L_{c}=84$ Mm, $L_{fp}=10$ Mm, $v_{A,c}=440$ km s$^{-1}$ and $v_{A,fp}=190$ km s$^{-1}$, we obtain: $P\sim590/m$ s, which matches well the resonant periods of the Alfv\'en waves (and hence the resulting periods for the longitudinal waves). 

\subsubsection{Photospheric magnetic field}\label{mag}

In Fig.~\ref{fig5} we plot the results of the parameter survey for different values of the photospheric magnetic field $B_{s,0}$. We test $B_{s,0}=2.3$ kG, 1.5 kG and 1 kG. Here the model consists of a 100 Mm length loop with an expansion factor of 1000 and a random photospheric driver. Having set in our model the value of the photospheric pressure to $p_{0}=2.09\times10^{5}$ dyn cm$^{-2}$, the equipartition magnetic field strength $B_{eq}=\sqrt{8\pi p_{0}}$ corresponds to $B_{s,0}= 2.3$ kG. In this case we use the more realistic random photospheric driver for the generation of Alfv\'en waves. From the panels of Fig.~\ref{fig5} we can see that the thermodynamical properties of the corona do not change considerably with the value of the photospheric magnetic field. For low amplitudes of the photospheric velocity field ($\langle v_{\phi}^{2}\rangle^{1/2}\lesssim1.1$ km s$^{-1}$) the photospheric magnetic energies, temperatures, densities, lengths and time spans of the corona are similar as shown by panels a, b, c, f, g and h. Also, the critical velocity above which the corona is unstable is roughly the same: $\langle v_{\phi}^{2}\rangle^{1/2}\sim2.2$ km s$^{-1}$. The a and b panels in Fig.~\ref{fig2} show that the nonlinear effects are also roughly similar, since these are rather independent on the background magnetic field. For photospheric velocities between 1.2 and 2.2 km s$^{-1}$, however, hotter and denser coronae are obtained for stronger photospheric magnetic field, as shown by panels a and b in Fig.~\ref{fig5}. For instance, for a photospheric velocity amplitude of $\sim2$ km s$^{-1}$ the coronal temperature and density are, respectively, $7.4\times10^{5}$ K and $2.5\times10^{8}$ cm$^{-3}$ for $B_{s,0}=$1 kG, and $1.1\times10^{6}$ K and $5.5\times10^{8}$ cm$^{-3}$ for $B_{s,0}=2.3$ kG. However, the plasma $\beta$ parameter is roughly equal for velocities $\langle v_{\phi}^{2}\rangle^{1/2}>2$ km s$^{-1}$ as seen in panel d. For lower values of the photospheric velocity field we obtain higher plasma $\beta$ the lower the value of the photospheric magnetic field. 

The disparities between the three cases can be understood by the different pressure distributions along the loop obtained by varying the photospheric magnetic field. Since the equipartition magnetic field corresponds to the case $B_{s,0}=2.3$ kG, the other cases have higher plasma $\beta$ at the same height in the photosphere and the chromosphere. As the amplitude of the torsional motions in the photosphere increases the magnetic pressure increases and pushes photospheric and chromospheric matter upwards. Since the magnetic pressure depends not only on the square of the torsional motions but also on the square of the longitudinal magnetic field the uplift effect is stronger for stronger photospheric magnetic field. Then, the gas pressure $p$ increases faster the higher the magnetic field. A higher coronal gas pressure implies a higher volumetric heating rate as shown by Eq.~\ref{hrtv}, and illustrated by the e panel of Fig.~\ref{fig5}, thus explaining the higher temperatures.

New observations with the Solar Optical Telescope on board of the Hinode satellite have shown patches of strong magnetic of 1 - 1.5 kG scattered uniformly all over the surface of the Sun, especially in Quiet Sun and Polar regions \citep{Tsuneta_2008ApJ...688.1374T}. Our results indicate that Alfv\'en waves that may be generated at the footpoints of loops spanning from these regions do not have energy limitations as far as the photospheric magnetic field is concerned. 

\subsubsection{Loop expansion}\label{expansion}

We now consider different expansions of the loop displaying area ratios of 1000, 600 and 500 between the corona and the photosphere. The loop is 100 Mm in length and has a photospheric magnetic field of 2.3 kG. The Alfv\'en waves are generated with the random photospheric driver. 

Fig.~\ref{fig6} displays the influence of the expansion of the loop in the thermodynamic properties of the corona. As the expansion of the loop is reduced the temperature and density in the corona are reduced as well, as shown by panels a and b, despite the same photospheric magnetic energy at the footpoints (panel c). For the loop with an expansion factor of 500 all the average temperatures in the corona are well below 1 MK (in this case the maximum achieved average temperature for a stable corona is $7.4\times10^{5}$ K, which appears to be an exception due to the large difference with the other cases). Consequently the plasma $\beta$ differs roughly by one order of magnitude in the corona between the loops with expansion factor 1000 and 500 (panel d). Together with the temperatures and densities of the corona, the critical velocities above which we have unstable coronae are reduced as well with lower loop expansions, as shown by panel h. While the critical velocity above which the corona collapses is $\langle v_{\phi}^{2}\rangle^{1/2}\sim2.2$ km s$^{-1}$ for a loop expansion of 1000, the corresponding critical velocities for loop expansions of 600 and 500 are $\sim1.7$ and $\sim1.3$ km s$^{-1}$ respectively. In the stable velocity range the lengths of the coronae are comparable, although there are 2 cases for the expansion factor of 500 for which the corona is barely stable and the loop lengths are reduced by half, namely, $\langle v_{\phi}^{2}\rangle^{1/2}\sim0.9$  and $\sim1$ km s$^{-1}$. We have also performed simulations with loop expansions of 300 but no stable corona was found.

As discussed in section \ref{model}, the expansion of magnetic flux tubes is a parameter which can greatly constrain the wave heating. Thin flux tubes allow large energy flows from Alfv\'en waves into the corona since they do not introduce any Alfv\'en velocity gradients (as long as the external temperature and internal temperature of the flux tubes do not differ), thus allowing cut-off free propagation of Alfv\'en waves \citep{Musielak_2007ApJ...659..650M, Verth_etal_aap_09}. When expansion is strong enough to introduce velocity gradients the spectrum of Alfv\'en waves that do not become evanescent is reduced \citep{Routh_2007SoPh..246..133R} , as discussed in section \ref{driver} \citep[caution should be taken however when applying the results stated in the later paper, as pointed out by][]{Verth_etal_aap_09}. However, in our case we have lack of heating for thinner flux tubes, while the strong expansion case of 1000 can easily achieve temperatures above 1 MK. 

A simple calculation can illustrate the dependence of mode conversion on the expansion of the loop. Assuming a typical Alv\'en speed of 10 km s$^{-1}$ in the upper photosphere/ lower chromosphere and a period of 100 s we obtain a wavelength of 1000 km for Alfv\'en waves propagating from the photosphere. Now, all the loops considered here display roughly the same initial expansion in the photosphere. We have an expansion of roughly 1000 km in the first 1000 km above the photosphere, thus matching the wavelength of the Alfv\'en wave. A slightly higher expansion has been reported by \citet{Jess_2009Sci...323.1582J} by comparing H$\alpha$ continuum with H$\alpha$ core images in observations of a bright point with the SOUP instrument of SST. Since this region is the region where the plasma $\beta$ is close to unity we have ideal conditions for mode conversion for the Alfv\'en waves. For instance, deformation of the wave shape and excitation of slow modes should take place \citep{Moore_1991ApJ...378..347M}. Since our 3 cases have roughly the same initial expansion, this implies that mode conversion taking place higher up in the chromosphere is very important for the Alfv\'en wave heating to be effective. 

The most important influence on the heating from the expansion of loops in our case has to do with the nonlinear effects. Higher expansions produce higher azimuthal velocities in the corona, which influence flows along the loop through the centrifugal force (cf. third term on the right hand side of equation \ref{momentums}). This in turn increases the density in the corona as seen in panel b of Fig.~\ref{fig6}, thus decreasing the Alfv\'en velocity.  Consequently the ratio $v_{\phi}/v_{A}$, which measures the nonlinear effects, increases. This is illustrated in panels a and c of Fig.~\ref{fig2}, where this ratio is plotted for loop expansions of 1000 and 500, for different amplitudes of the photospheric velocity field. From basic concepts it can be easily shown that 
\begin{equation}
\frac{v_{\phi}}{v_{A}}\propto A\rho^{1/4},
\end{equation}  
where $A$ is the area expansion coefficient. As seen in Paper 1, $v_{\phi}/v_{s}$ becomes high ubiquitously and sporadically in the corona. When this happens nonlinear effects are large and energy is transfered from the Alfv\'en mode to the longitudinal slow and fast modes, which steepen into shocks and heat the plasma (see section \ref{dissipation}). Hence, the reason for the absence of heating in our thin loops is not the lack of energy flux or larger radiative losses but a lack of dissipation of the waves. 

\subsubsection{Loop length}\label{length}

Observations of coronal loops and posterior analytical studies based on the energy balance in these closed magnetic field structures have shown that the coronal temperature in uniformly heated loops increases with the  length of the corona \citep{Rosner_1978ApJ...220..643R, Serio_1981ApJ...243..288S}. In order to study the effect of length on Alfv\'en wave heating here we consider coronal loops with lengths of 100, 80 and 60 Mm. The loops have the same expansion factor, namely 1000, and a photospheric magnetic field of 2.3 kG. The waves are generated with the random photospheric driver. 

Fig.~\ref{fig7} shows the effect of loop length in the creation of the corona from Alfv\'en wave heating. The results in this case are similar to the previous case in which the expansion of the loop is considered. We can see that as the length of the loop is reduced the heating of the loop is dramatically reduced (panel a), despite the similar magnetic energy input at the footpoints (panel c). Already for the 80 Mm length loop we cannot obtain a corona with an average temperature above 1 MK. Densities are also reduced and therefore also the plasma $\beta$ (panels b and d). The sensitivity of the coronae to the loop length can also be seen in panels f and h, where we can see that the range of photospheric velocities over which the coronae can be created and maintained is considerably reduced with respect to all the previous cases. The critical velocities above which the coronae collapse are  $\langle v_{\phi}^{2}\rangle^{1/2}\sim1.1$  and $\sim0.6$ km s$^{-1}$, respectively for 80 and 60 Mm lengths. For photospheric velocity ranges in which the corona is stable we can see from panel f that the lengths of the corona vary considerably. For $L=80$ Mm the lengths vary between 25 and 50 Mm. For $L=60$ Mm we have only 2 stable cases with lengths between 10 and 20 Mm. 

The lack of wave heating can be understood partly by considering the number of shocks at any time heating the corona. In section \ref{driver} we showed that long period waves produce hot coronae while short period waves account for their stability by heating them uniformly. The most efficient period range for the heating was found to be between 100 and 150 s for the longitudinal waves, thus 200 and 300 s for the Alfv\'en waves. For a photospheric velocity of $\sim1$ km s$^{-1}$ typical Alfv\'en velocities in the corona are $\sim$ 400 and $\sim$ 300 km s$^{-1}$ for loop lengths of 80 and 60 Mm respectively (as derived from panel b). Taking a period of 200 s, this gives wavelengths of 80 Mm and 60 Mm for the Alfv\'en waves respectively, which is just the length of the loops. We see that the wave has barely the distance to propagate one wavelength before reaching the other footpoint, where the dense sub-photospheric regions will not allow an easy reflection. Taking a roughly constant sound speed in the corona it is easy to see that the necessary distance for shock formation will be at least on the order of a wavelength (in the case of a high velocity amplitude). Since both footpoints generate Alfv\'en waves we then may have a maximum of 4 shocks propagating in the corona at any time (neglecting all other dissipative mechanisms such as wave-to-wave interaction), and may not easily reflect in the transition regions due to their long wavelengths. The same reasoning applied to a 100 Mm length loop gives roughly 12 shocks propagating at any time. Furthermore, in the later case shocks can easily reflect at the transition region and get trapped in the corona, thus increasing the dissipation in the corona. 

Fig.~\ref{fig2} shows that nonlinear effects are considerably reduced for shorter loop lengths. The highest value of the ratio $v_{\phi}/v_{A}$ for stable coronae is about 0.1, which is not enough for mode conversion to be efficient. The main reason for the lack of hot coronae in short loops is the same reason for their instability.  Due to the shorter length the effective gravity is significantly reduced and the height of the spicules increases considerably as shown in panel f of Fig.~\ref{fig7}. As the amplitudes of the waves increase large amounts of cool matter are input into the loop. Spicules have then a stronger cooling effect the shorter the loop length. 

\section{Discussion and conclusions}\label{four}

Alfv\'en waves constitute one of the main candidates for heating the solar corona. Theoretically easy to be generated from convective turbulent motions in the photosphere it has been shown that these waves are able to carry sufficient amounts of energy in order to balance energy losses from radiation and conduction and heat the corona up to the observed few million degrees \citep{Uchida_1974SoPh...35..451U, Wentzel_1974SoPh...39..129W, Hollweg_1982SoPh...75...35H, Poedts_1989SoPh..123...83P, Ruderman_1997AA...320..305R, Kudoh_1999ApJ...514..493K}. Finding a suitable dissipation mechanism is a very difficult task, which has spawned a large active research community in solar physics. In this work we have considered a model in which mode conversion acts as the dissipative mechanism. More precisely, through nonlinear effects based on density fluctuations and wave-to-wave interaction in the chromosphere and corona the Alfv\'en mode is able to transfer some part of its energy to the fast and slow modes, which steepen into shocks and heat the plasma. 

We have conducted a parameter survey in which the effect of geometrical quantities such as the loop expansion and the loop's length on the efficiency of Alfv\'en wave heating is studied. We further investigate the effect on the heating of other physical parameters such as the photospheric magnetic field and the generation of Alfv\'en waves with a monochromatic and white noise spectrum in the photosphere.

\begin{figure}
\epsscale{0.7}
\plotone{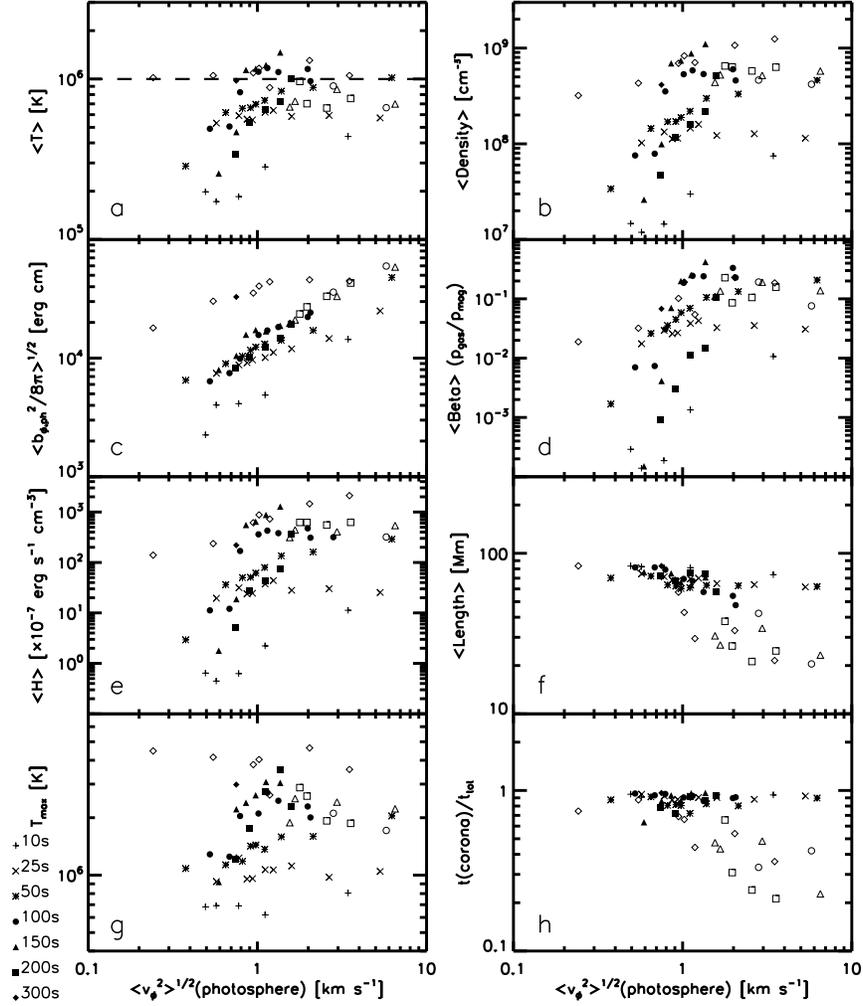}
\caption{Results of Alfv\'en wave heating from monochromatic waves. The (longitudinal) waves have periods of 10, 25, 50, 100, 150, 200 and 300 s (corresponding symbols are written in the bottom left part of the figure). Average quantities are plotted with respect to the rms azimuthal velocity amplitude at the photosphere. Panels a, b d and e: temperature, density, plasma $\beta$ and volumetric heating rate in the corona, respectively. The volumetric heating rate is calculated according to Eq.~\ref{hrtv}. Panel c: rms azimuthal magnetic energy in the photosphere.  Panel g: maximum attained temperatures. Panels f and h: length and time span of the corona (with respect to total simulation time), respectively. Blank symbols denote unstable coronae (see \S \ref{collapse}). \label{fig3}}
\end{figure}

\begin{figure}
\epsscale{1.0}
\plottwo{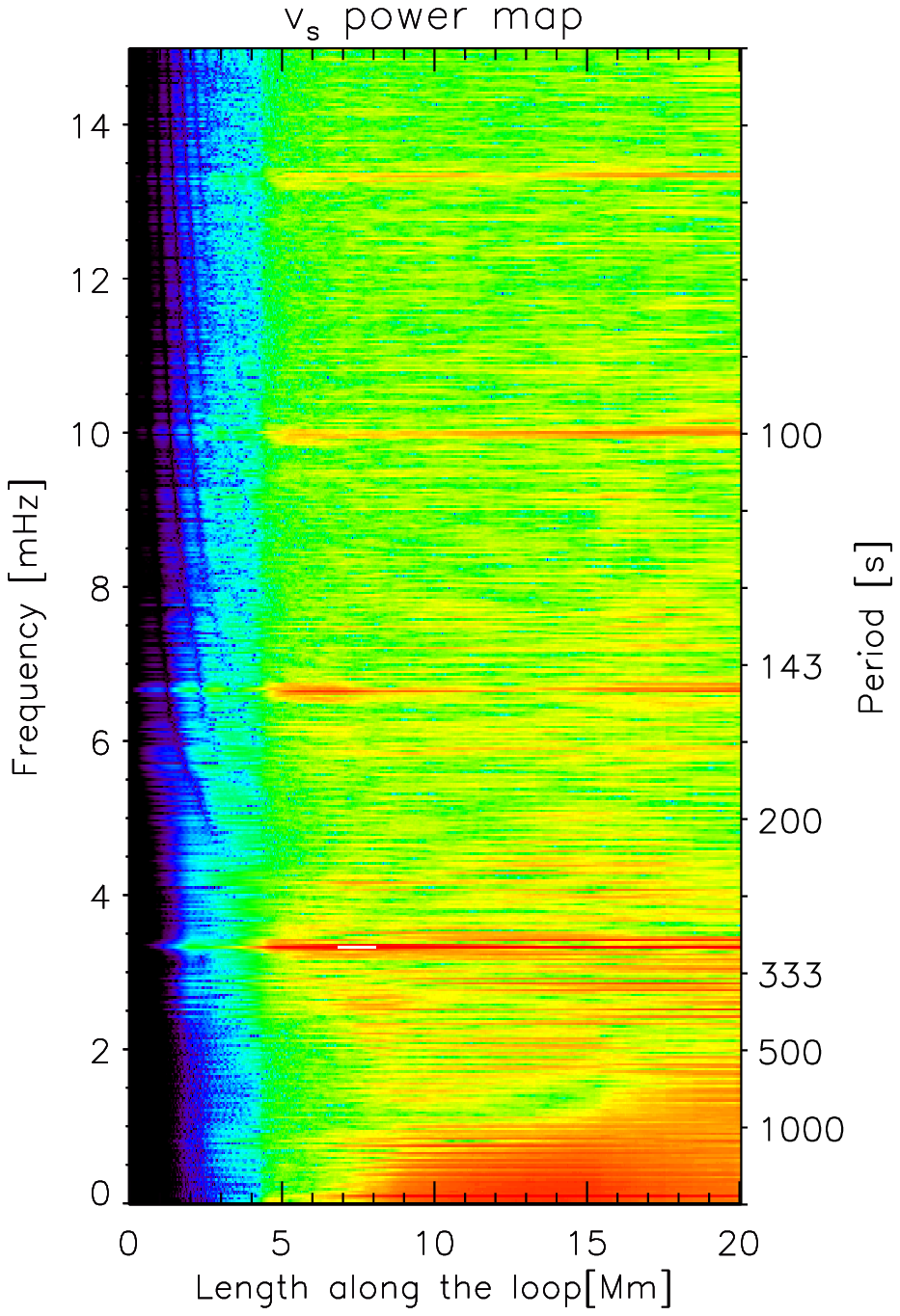}{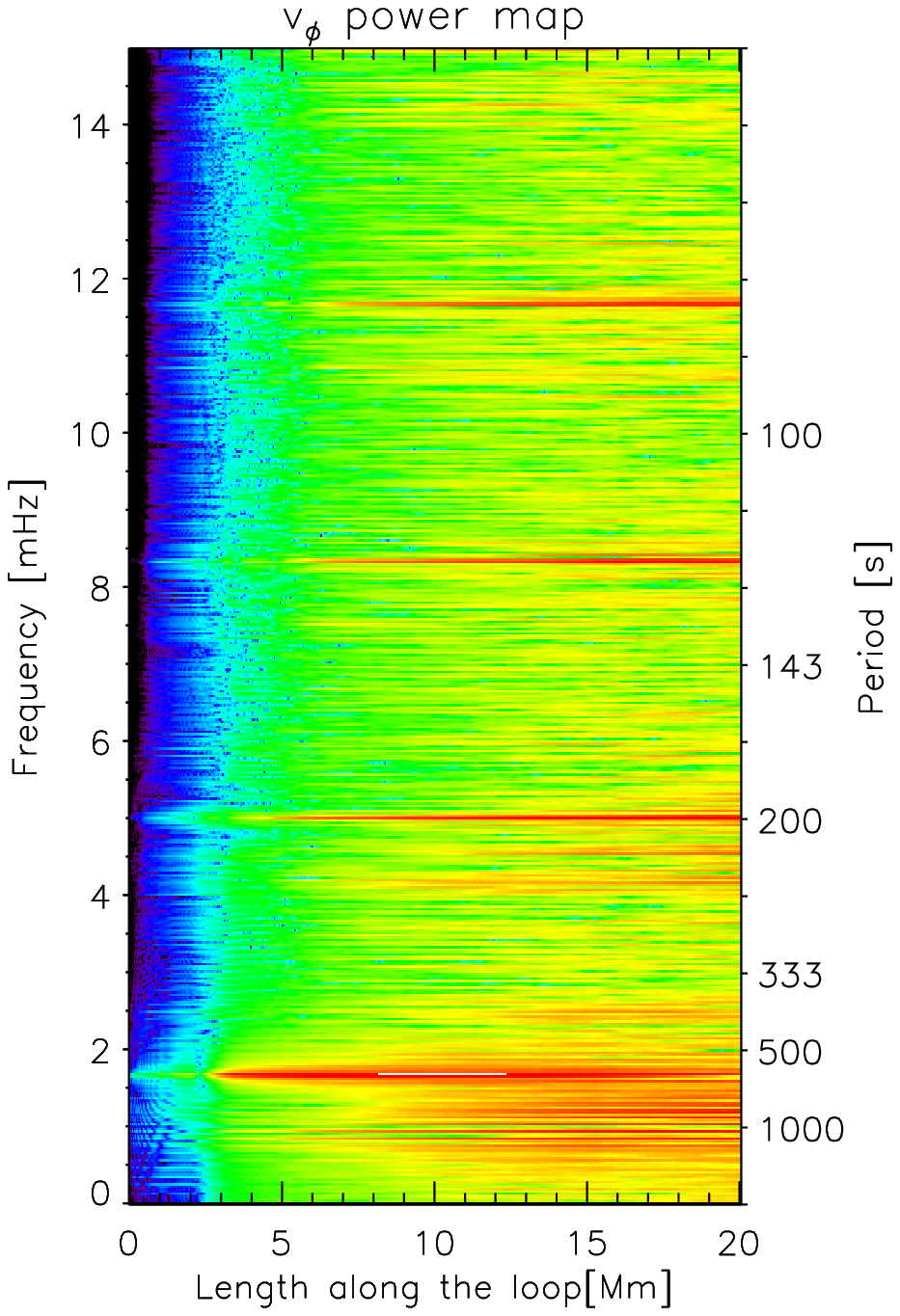}
\caption{Map of the power spectrum in logarithmic scale of the longitudinal velocity component $v_{s}$ (left) and azimuthal velocity component $v_{\phi}$ (right) for the first 20 Mm along a loop where 300 s monochromatic longitudinal waves are generated (from mode conversion of 600 s Alfv\'en waves). \label{fig4}}
\end{figure}

\begin{figure}
\epsscale{0.7}
\plotone{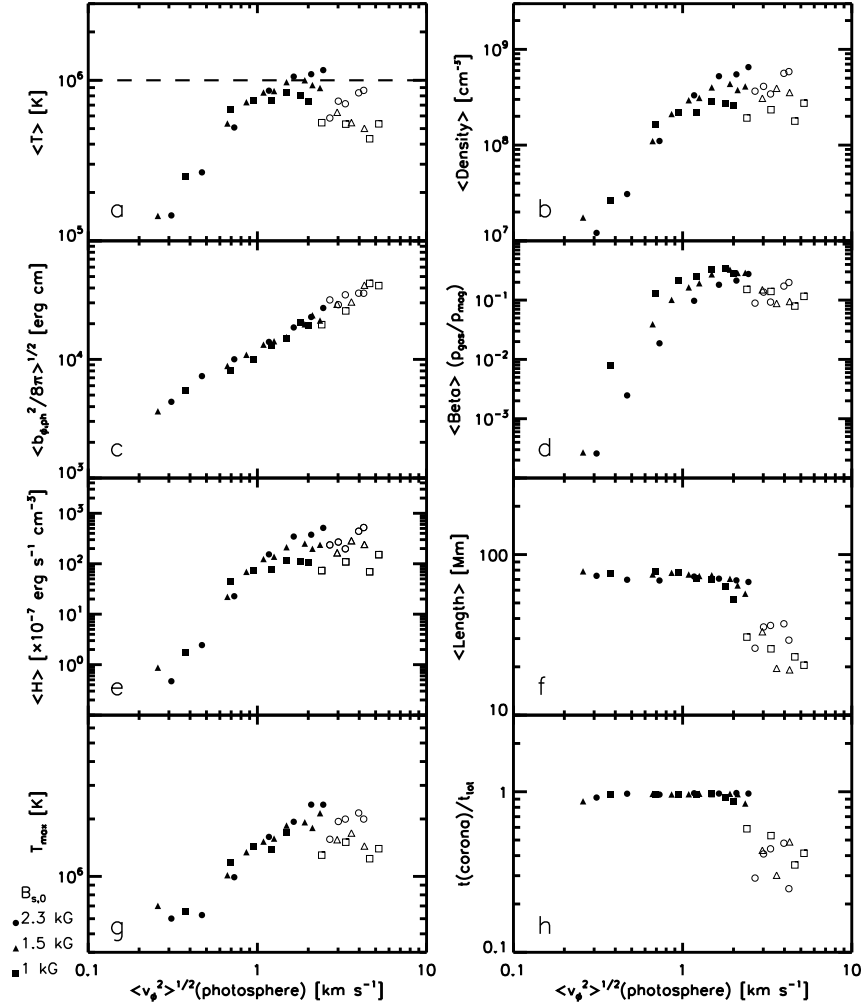}
\caption{Results of Alfv\'en wave heating with different photospheric magnetic fields. Fields of 2.3 kG (circles), 1.5 kG (triangles) and 1 kG (squares) are tested. See legend of Fig.~\ref{fig3} for explanation of the panels.    \label{fig5}}
\end{figure}

\begin{figure}
\epsscale{0.7}
\plotone{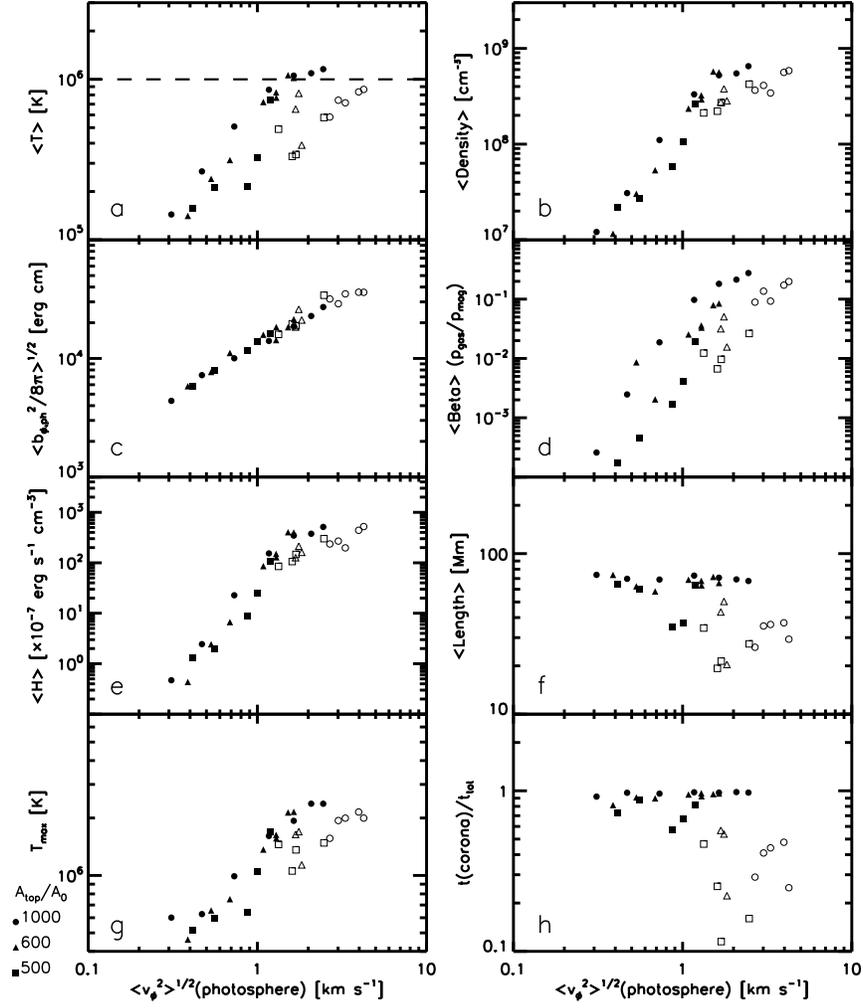}
\caption{Results of Alfv\'en wave heating with different loop expansion factors $A_{top}/A_{0}$, where $A_{s}$ is the cross-section area of the loop and $s=0, top$ correspond to the photosphere, apex of the loop (corona), respectively. Expansion of 1000 (circles), 600 (triangles) and 500 (squares) are tested. See legend of Fig.~\ref{fig3} for explanation of the panels.   \label{fig6}}
\end{figure}

\begin{figure}
\epsscale{0.7}
\plotone{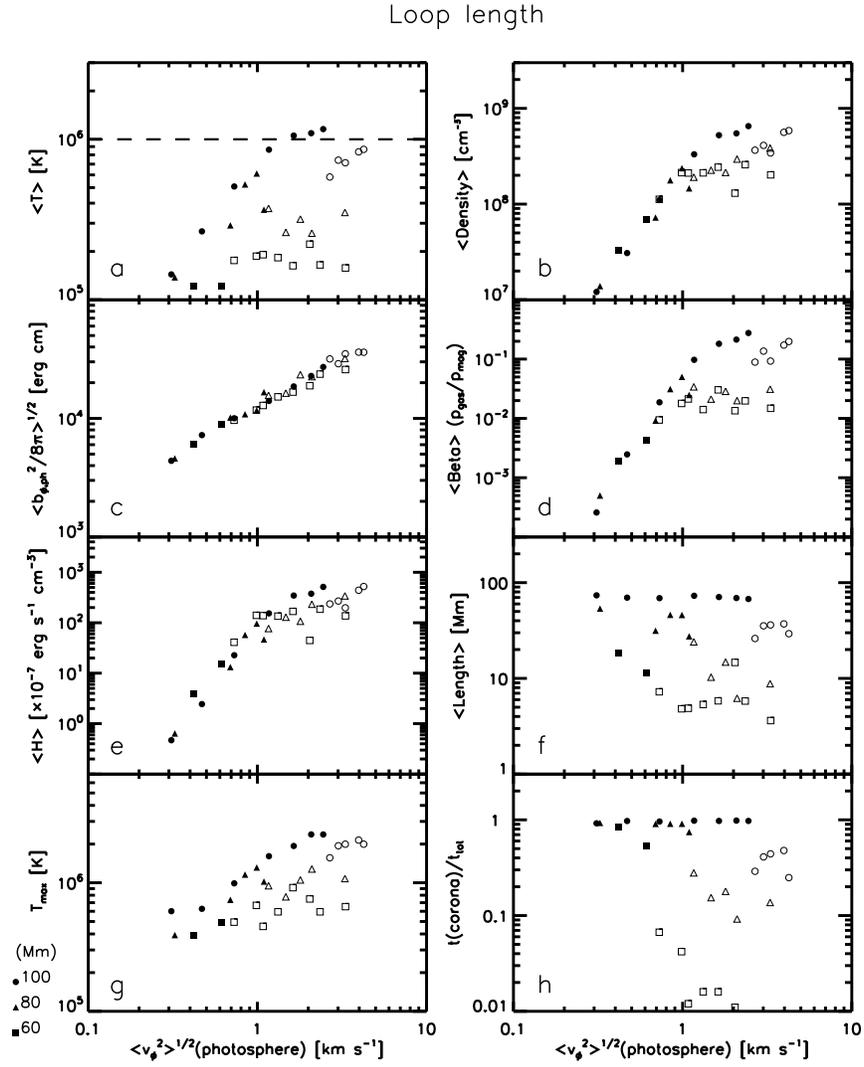}
\caption{Results of Alfv\'en wave heating with different loop lengths. Lengths of 100 Mm (circles), 80 Mm (triangles) and 60 Mm (squares) are tested. See legend of Fig.~\ref{fig3} for explanation of the panels.   \label{fig7}}
\end{figure}

\begin{figure}
\epsscale{1.1}
\plottwo{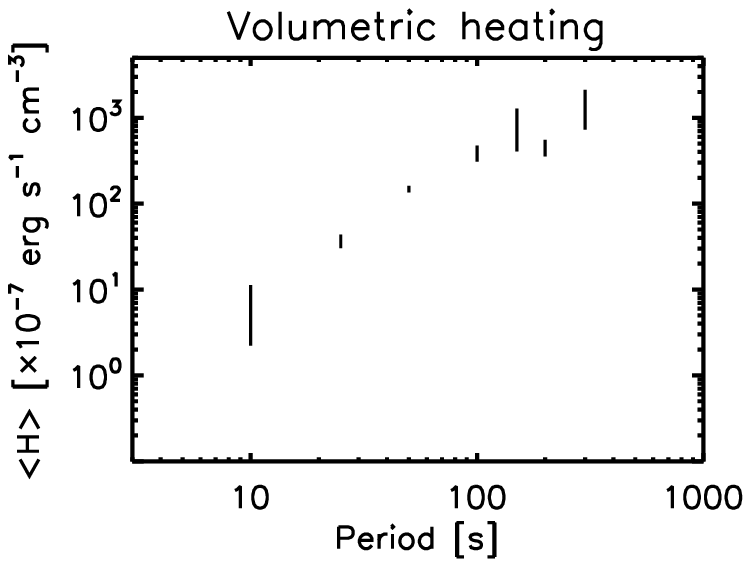}{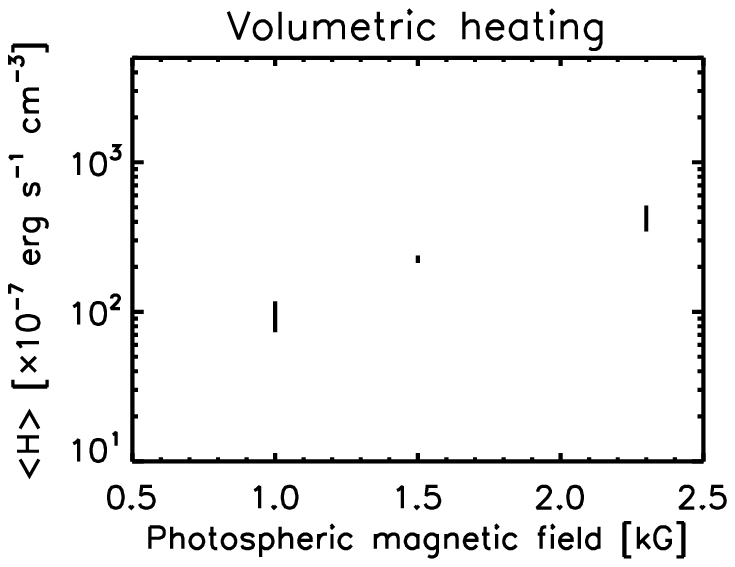}\\
\epsscale{1.1}
\plottwo{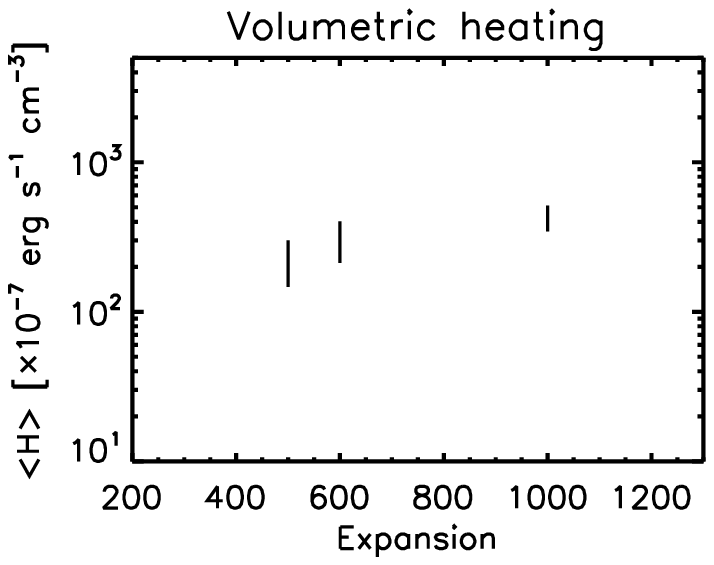}{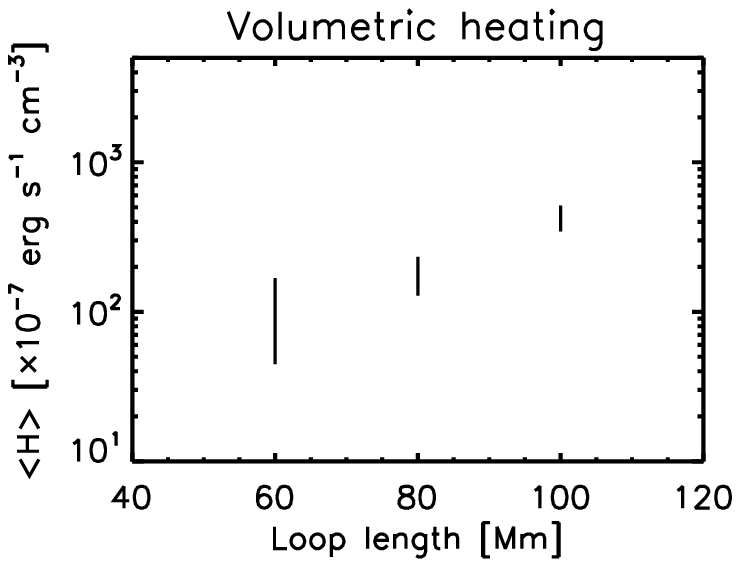}
\caption{Volumetric heating rate with respect to the considered parameters of the loop: period of the monochromatic longitudinal waves issuing from mode conversion of Alfv\'en waves (top left panel), photospheric magnetic field (top right panel), expansion of the loop (bottom left panel), and loop length (bottom right panel). The volumetric heating rate for each case corresponds to its extremum values in the range between $\langle v_{\phi}^{2}\rangle^{1/2}\sim1.3$  and 2.3 km s$^{-1}$ found in the e panels of Figs.~\ref{fig3}, \ref{fig5}, \ref{fig6} and \ref{fig7}. \label{fig8}}
\end{figure}

In Fig.~\ref{fig8} we plot the volumetric heating rate with respect to the considered parameters of the loop: period of the monochromatic longitudinal waves issuing from mode conversion of Alfv\'en waves (top left panel), photospheric magnetic field (top right panel), expansion of the loop (bottom left panel), and loop length (bottom right panel). The volumetric heating rate for each case corresponds to its extremum values in the range between $\langle v_{\phi}^{2}\rangle^{1/2}\sim1.3$  and 2.3 km s$^{-1}$ found in the e panels of Figs.~\ref{fig3}, \ref{fig5}, \ref{fig6} and \ref{fig7}, which is the most efficient amplitude range for coronal heating we have found. We can see that the volumetric heating rate is an increasing function for all the considered parameters in the chosen azimuthal photospheric velocity interval. This figure summarizes well the results of our parameter survey. As the amplitudes of the twists in the photosphere generating the Alfv\'en waves increase the momentum and energy flux of the waves increase. We have found that nonlinear effects generally increase as well, and mode conversion happens not only in the chromosphere but ubiquitously in the corona, thus heating the plasma. Suitable conditions for Alfv\'en wave heating are thus strongly dependent on the nonlinear effects. The importance over nonlinearity favors thick and long loops (bottom left and right panels), and strong photospheric magnetic fields (top right panel). The dependence on the later is however not crucial and flux tubes with 1 kG field concentrations at their footpoints may be heated efficiently by Alfv\'en waves. On the other hand, the expansion and the length of the loop are crucial parameters for the heating. As we can see in Fig.~\ref{fig8}, the volumetric heating rate in the bottom left and right panels does not vary much respect to the loop expansion and the length, which shows the high sensitivity of the thermodynamics in the loop on the heating. Nonlinearity is directly proportional to the loop expansion. Furthermore, the strong expansion of loops from the photosphere to the transition region happens in a height comparable to the wavelength of the Alfv\'en wave, which leads to the deformation of the Alfv\'en wave and excitation of slow modes \citep{Moore_1991ApJ...378..347M}. Since the plasma $\beta$ parameter in that region is close to unity, further mode conversion is expected. Alfv\'en waves have long wavelengths and in order for them to convert into longitudinal waves which steep into shocks they need to propagate relatively long distances. Loops with lengths lower than 80 Mm cannot be heated by Alfv\'en waves in the present model. Caution must be taken however when drawing conclusions from these results. When calculating the lengths of loops in observations what is actually calculated is the length of the corona, which is the visible part of loops in EUV or X-ray images. In our model, the lengths of the coronae with average temperatures of $\sim1$ MK we obtain with Alfv\'en wave heating oscillate between 50 and 80 Mm with a mean around 70 Mm. Furthermore, we have defined a stable corona, as a corona which can be maintained over roughly 5.7 hours. However, observations show that coronal loops are dynamical entities which exhibit heating and cooling processes constantly. Likewise, the loops with a collapsing corona obtained here exhibit heating events with nanoflare-like temperatures constantly, which can match the observed bursty X-ray intensity profiles and the statistics displaying power law distribution of the heating events (cf. Paper 1). Also, a loop which is visible in soft X-rays over a large period of time may in fact result from the effect of different threads in the loop being heated at different times, thus giving the impression of a steady and uniformly heated loop \citep[as discussed in][]{Patsourakos_2009ApJ...696..760P}. In the simple model we have considered we have assumed that the heating acts uniformly along the radial direction, thus disabling the possibility of many threads being heated at different times. We thus cannot completely reject the possibility of Alfv\'en waves heating the coronae of short and thin loops. This idea should however be tested in an extended (at least 2.5 dimensional) version of this model.

All coronae show a range of velocities for which they are stable throughout the simulation and for which their lengths are constant. However, when the photospheric velocity field exceeds a critical value the corona collapses. The critical velocity is dependent on the parameters of the model. In this regime wave heating can no longer account for the large cool mass flux from the increasing wave amplitudes and collapses. Generally, the collapse is not due to a local increase in the corona of the radiative losses from the formation of a cool condensation as in the case of catastrophic cooling \citep{Antiochos_1999ApJ...512..985A}, and does not exhibit ``limit cycles" \citep{Muller_2003AA...411..605M}. The later are characteristic of the catastrophic cooling mechanism which is proposed as an explanation for coronal rain. Further investigation in that direction is however required and will be the subject of a future paper. 

All of the obtained coronae are uniformly heated irrespective of the parameters in the model. 
This is a characteristic of Alfv\'en wave heating. The distribution of the heating in coronal loops as derived from observations is a matter of debate. Indeed, different methods for the analysis of observational data may lead to different results. A famous example of this fact is the analysis of the data set from soft X-ray observations with the \textit{Yohkoh}/SXT instrument. An interpretation of the heating distribution of the observed loops in terms of apex concentrated heating \citep{Reale_2002ApJ...580..566R}, footpoint concentrated heating \citep{Aschwanden_2001ApJ...559L.171A} and uniform heating \citep{Priest_1998Natur.393..545P} has been given. There seems to be however more observational evidence of coronal loops being heated towards the footpoints in active regions \citep{Aschwanden_2001ApJ...560.1035A, Aschwanden_2000ApJ...541.1059A}. Further evidence of this fact has been found by \citet{Hara_2008ApJ...678L..67H} using the \textit{Hinode}/EIS instrument, in which active region loops are shown to exhibit upflow motions and enhanced nonthermal velocities in the hot lines of Fe XIV 274 and Fe XV 284. Possible unresolved high-speed upflows were also found. In Paper 1 we found that footpoint or uniform nanoflare heating exhibit hot upflows, thus fitting in the observational scenario of active regions, while Alfv\'en wave heating was found to exhibit hot downflows, which may fit in the observational scenario of quiet Sun regions  \citep{Chae_1998ApJS..114..151C, Brosius_2007ApJ...656L..41B}. The uniform heating from Alfv\'en waves further supports this conclusion, since it is a characteristic of loops more generally found in quiet Sun regions. Active region loops exhibit low expansion factors due to the high magnetic field filling factor in those regions. In this chapter we have found that Alfv\'en wave heating is effective only in thick loops (with area expansions between photosphere and corona higher than 600), further emphasizing our conclusion that active regions may not be heated by Alfv\'en waves. These results seem to point to an important role of Alfv\'en wave heating in Quiet Sun regions, where loops are often long, expand more than in active regions, and kG (or higher) bright points are ubiquitous. 

Another important result we have found is that both long (above 100 s) and short period (below 50 s) waves (where the period is the resulting period of the longitudinal modes from mode conversion; the period of the Alfv\'en waves is twice the stated period) play an important role in the heating of the loops. Long period waves produce very hot heating events in the corona and thus increase the average temperature of the corona due to the strong shocks they produce. Short period waves are responsible for keeping the corona, thus making it stable through uniform heating from the numerous weak shocks they produce. We have therefore a compromise between long and short period waves leading to efficient Alfv\'en wave heating. The 300 s period waves are found to have considerable power to heat the corona even with low photospheric velocity amplitudes of $\sim0.2$ km s$^{-1}$. For these waves resonances seem to be triggered in the loop, which allow higher transmission from the waves into the corona and subsequent mode trapping leading to efficient dissipation. This resonant damping mechanism is however not observed for random wave generation (white noise spectrum) or with other geometric parameters for the loop. A thorough study of these coronal resonances for Alfv\'en waves will be the subject of a future paper.

\acknowledgments

P. A. would l ike to thank M. Carlsson, V. Hansteen, L. Heggland, E. Leer, R. Erd\'elyi, T. Matsumoto, K. Ichimoto, T. Suzuki and H. Isobe for many fruitful discussions. P. A. would also like to acknowledge S. F. Chen for patient encouragement. This work was supported by the Grant-in-Aid for the Global COE Program ``The Next Generation of Physics, Span from Universality and Emergence'' from the Ministry of Education, Culture, Sports, Science and Technology ( MEXT ) of Japan, by a Grant from the International Astronomical Union, and by a Grant-in-Aid for Creative Scientific Research, ``The Basic Study of Space Weather Prediction'' (17GS0208; Head Investigator: K. Shibata), from the Ministry of Education, Science, Sports, Technology, and Culture of Japan. The numerical calculations were carried out on Altix3700 BX2 at YITP in Kyoto University.

\bibliographystyle{aa}
\bibliography{aamnemonic,patbib}  

\clearpage

\end{document}